\documentclass[12pt]{article}

\def\ltap{\raisebox{-.4ex}{\rlap{$\,\sim\,$}} \raisebox{.4ex}{$\,<\,$}}
\def\gtap{\raisebox{-.4ex}{\rlap{$\,\sim\,$}} \raisebox{.4ex}{$\,>\,$}}
\newcommand\as{\alpha_{\mathrm{S}}}
\def\beq{\begin{equation}}
\def\eeq{\end{equation}}
\def\beeq{\begin{eqnarray}}
\def\eeeq{\end{eqnarray}}
\def\bom#1{{\mbox{\boldmath $#1$}}}
\def\to{\rightarrow}

\def\nn{\nonumber}

\begin{document}

\begin{titlepage}
\renewcommand{\thefootnote}{\fnsymbol{footnote}}
\begin{flushright}
     CERN--TH/2000-073 
\\ hep-ph/0005233 
     \end{flushright}
\par \vspace{10mm}

\begin{center}
{\Large \bf Aspects of QCD, from the Tevatron to the LHC~\footnote{To appear in the
Proceedings of the Workshop {\it Physics at TeV Colliders}, Les Houches, 
France 8--18 June 1999.}\footnote{This work was 
supported in part 
by the EU Fourth Framework Programme ``Training and Mobility of Researchers'', 
Network ``Quantum Chromodynamics and the Deep Structure of
Elementary Particles'', contract FMRX--CT98--0194 (DG 12 -- MIHT).}}
\end{center}

\par \vspace{2mm}
\begin{center}
{\bf Stefano Catani}~\footnote{On leave of absence from INFN,
Sezione di Firenze, Florence, Italy.}\\

\vspace{5mm}

{Theory Division, CERN, CH 1211 Geneva 23, Switzerland} 

\vspace{5mm}

\end{center}

\par \vspace{2mm}
\begin{center} {\large \bf Abstract} \end{center}
\begin{quote}
\pretolerance 10000

This contribution presents a selection of the topics 
(parton densities, fixed-order calculations, parton showers, soft-gluon
resummation) discussed in my introductory lectures at the Workshop and 
includes a pedagogical overview of the corresponding theoretical
tools.

\end{quote}

\vspace*{\fill}
\begin{flushleft}
     CERN--TH/2000-073 \\     February 2000 
\end{flushleft}
\end{titlepage}

\renewcommand{\thefootnote}{\fnsymbol{footnote}}

\section{Introduction}
\label{intro}

The production cross sections for all the processes at hadron-collider
experiments are controlled by strong interaction physics and, hence,
by its underlying field theory, QCD (see recent overviews in 
Refs.~[\ref{eps99}, \ref{lp99webber}, \ref{ichep98}, \ref{lp97}]).
Studies of QCD at the Tevatron and the LHC
have two main purposes [\ref{proceeding}, \ref{proctev}, \ref{proclhc}]. 
First, they are important to test the predictions of
QCD, to measure its fundamental parameters (e.g. the strong coupling $\as$) and
to extract quantitative information on its non-perturbative dynamics (e.g. the
distribution of partons in the proton). Second, they are relevant to a precise 
estimate of the background to other Standard Model processes and to signals of
new physics.

This contribution is not a comprehensive review of QCD at high-energy hadron
colliders. It is based on a selection of the topics presented in my
introductory lectures at this Workshop. The selection highlights
the QCD subjects that were most discussed during the Workshop 
and includes a pedagogical overview of some of the corresponding theoretical
tools. 

After the introduction of the general theoretical framework, 
I~summarize in Sect.~\ref{secpdf} 
the present knowledge on the parton densities and its impact on
QCD predictions for hard-scattering processes at the Tevatron and the LHC.
In Sect.~\ref{secgdensity}, I then discuss some issues related to processes
that are sensitive to the gluon density and, hence, to its 
determination. Section~\ref{secpxs} presents a dictionary of 
different approaches (fixed-order expansions, resummed calculations, 
parton showers) to perturbative QCD calculations. 
The dictionary continues in Sect.~\ref{secsoftg}, where I review soft-gluon
resummation and discuss some recent phenomenological applications of threshold
resummation to hadron collisions.

The QCD framework to describe any inclusive hard-scattering process,
\beq
\label{hardpro}
h_1(p_1) + h_2(p_2) \to H(Q,\{ \dots \}) + X \;\;,
\eeq
in hadron--hadron collisions is based on perturbation theory and on the 
factorization theorem of mass singularities. The corresponding cross section
is computed by using the factorization formula [\ref{factform}]
\beeq
\sigma(p_1,p_2;Q, \{ \dots \} ) \!\!\!&=&\!\!\! 
\sum_{a,b} \int_{x_{\rm min}}^1 dx_1 \, dx_2 \,f_{a/h_1}(x_1, \mu_F^2)
\, f_{b/h_2}(x_2, \mu_F^2) \; 
{\hat \sigma}_{ab}(x_1p_1,x_2p_2;Q, \{ \dots \}; \mu_F^2)
\nn \\
\label{factfor}
\!\!\!&+&\!\!\! {\cal O}\left( (\Lambda_{QCD}/Q )^p \right) \;\;.
\eeeq

The colliding hadrons $h_1$ and $h_2$ have momenta $p_1$ and $p_2$, $H$ denotes
the triggered hard probe (vector bosons, jets, heavy quarks, Higgs bosons,
SUSY particles and so on) and $X$ stands for 
any unobserved particle produced by the collision. The typical scale $Q$
of the scattering process is set by the invariant mass or the transverse
momentum of the hard probe, and the notation $\{ \dots \}$ stands for any other
relevant scale and kinematic variable of the process. For instance,
in the case of $W$ production we have $Q=M_W$ and $\{ \dots \}= \{ Q_{\perp},
y, \dots \}$, where $M_W, Q_{\perp}$ and $y$ are the mass of the vector boson, 
its transverse momentum and its rapidity, respectively.

The factorization formula (\ref{factfor}) involves the convolution
of the partonic cross sections ${\hat \sigma}_{ab}$ (where $a,b=q,{\bar q},g)$ 
and the parton distributions $f_{a/h}(x, \mu_F^2)$ 
of the colliding hadrons. If the hard probe $H$ is a hadron or a photon,
the factorization formula has to include an additional convolution with the
corresponding parton fragmentation function $d_{a/H}(z, \mu_F^2)$.

The term ${\cal O}\left( (\Lambda_{QCD}/Q )^p \right)$ on the right-hand side 
of Eq.~(\ref{factfor}) generically denotes non-perturbative contributions
(hadronization effects, multiparton interactions, contributions of the soft
underlying event, and so on). Provided the hard-scattering process 
(\ref{hardpro}) is sufficiently inclusive\footnote{ 
More precisely, it has to be defined
in an infrared- and collinear-safe manner.}, 
${\hat \sigma}_{ab}$ is computable as a power series expansion in $\as(Q^2)$
and the non-perturbative contributions are (small) power-suppressed corrections
(i.e. the power $p$ is positive) as long as the hard-scattering scale
$Q$ is larger than few hundred MeV, the typical size of the QCD scale 
$\Lambda_{QCD}$.

The parton densities $f_{a/h}(x, \mu_F^2)$ are phenomenological distributions 
that describe how partons are bounded in the colliding hadrons. Although they 
are not calculable in QCD perturbation theory, the parton densities
are universal (process-independent) quantities. The scale $\mu_F$ is a
factorization scale introduced in Eq.~(\ref{factfor}) to separate the
bound-state effects from the perturbative interactions of the partons.
The physical cross section $\sigma(p_1,p_2;Q, \{ \dots \} )$ does not depend on
this arbitrary scale, but parton densities and partonic cross sections 
separately depend on $\mu_F$. In particular, higher-order contributions to 
${\hat \sigma}_{ab}(x_1p_1,x_2p_2;Q, \{ \dots \}; \mu_F^2)$ contain corrections
of relative order $(\as(Q^2) \ln Q^2/\mu_F^2)^n$. If $\mu_F$ is very different
from $Q$, these corrections become large and spoil the reliability of the
perturbative expansion. Thus, in practical applications of the factorization
formula (\ref{factfor}), the scale $\mu_F$ is set approximately equal to
the hard scale $Q$ and variations of $\mu_F$ around this central value
are used to estimate the uncertainty of the perturbative expansion.

\begin{figure}
  \centerline{
    \setlength{\unitlength}{1cm}
    \begin{picture}(0,6)
       \put(0,0){\includegraphics{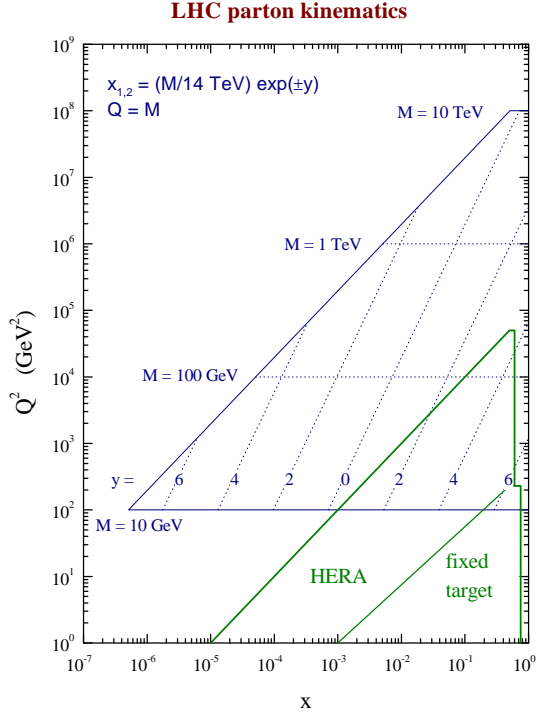}}
    \end{picture}}
\vspace*{3.0cm}    
\caption{The $(x,Q^2)$ plane of the parton kinematics for the production of
a heavy system of invariant mass $M$ and rapidity $y$ at LHC, HERA and
fixed-target experiments.
\label{lhckin}}
\end{figure}

The lower limit $x_{\rm min}$ of the integrations over the parton momentum
fractions $x_1$ and $x_2$, as well as the values of $x_1$ and $x_2$ that
dominate the convolution integral in Eq.~(\ref{factfor}), are controlled
by the kinematics of the hard-scattering process. Typically we have
$x_{\rm min} \gtap Q^2/S$, where $S=(p_1+p_2)^2$ is the square of the
centre-of-mass energy of the collision. If the hard probe is a state of
invariant mass $M$ and rapidity $y$, the dominant values of the momentum
fractions are $x_{1,2} \sim (M e^{\pm y})/{\sqrt S}$
(see Fig.~\ref{lhckin}). 
Thus varying $M$ and
$y$ at fixed ${\sqrt S}$, we are sensitive to partons with different
momentum fractions. Increasing ${\sqrt S}$ the parton densities are probed in
a kinematic range that extends towards larger values of $Q$ and smaller values
of $x_{1,2}$.

\section{Parton densities}
\label{secpdf}

The parton densities are an essential ingredient to study hard-scattering
collisions. Once the partonic cross sections have been perturbatively
computed, cross section measurements can be used to determine the parton
densities. Then, they can in turn be used to predict cross sections
for other hard-scattering processes.

The dependence of the parton densities\footnote{In the following the parton
densities of the proton $f_{a/p}$ are simply denoted by $f_{a}$ and
those of the antiproton are obtained by using charge-conjugation invariance,
i.e. $f_{a/{\bar p}}= f_{{\bar a}/p} = f_{{\bar a}}$.}
$f_{a}(x, \mu^2)$ on the momentum fraction $x$ and their absolute value
at any fixed scale $\mu$
are not computable in perturbation theory. However, the scale dependence is
perturbatively controlled by the DGLAP evolution equation~[\ref{DGLAP}] 
\beq
\label{evequa}
\frac{d \,f_{a}(x, \mu^2)}{d \ln \mu^2} = 
\sum_{b} \int_{x}^1 \frac{dz}{z} \, P_{ab}(\as(\mu^2), z) \,f_{a}(x/z, \mu^2)
\;\;.
\eeq
The kernels $P_{ab}(\as, z)$ are the Altarelli--Parisi (AP) splitting functions.
As the partonic cross sections in Eq.~(\ref{factfor}), the AP splitting 
functions can be computed as a power series expansion in $\as$:
\beq
\label{apexp}
P_{ab}(\as, z) = \as P_{ab}^{(LO)}(z) + \as^2 P_{ab}^{(NLO)}(z)
+ \as^3 P_{ab}^{(NNLO)}(z) + {\cal O}(\as^4) \;\;.
\eeq
The leading order (LO) and next-to-leading order (NLO) terms 
$P_{ab}^{(LO)}(z)$ and $P_{ab}^{(NLO)}(z)$ in the expansion 
are known [\ref{NLOAP}]. These first two terms are used
in most of the QCD studies.
Having determined $f_{a}(x, Q_0^2)$ at a given input scale $\mu = Q_0$, the
evolution equation (\ref{evequa}) can be used to compute the parton densities
at different perturbative scales $\mu$ and larger values of $x$.

The parton densities are determined by performing global fits 
[\ref{pdffit}] to data from deep-inelastic scattering (DIS), Drell--Yan (DY),
prompt-photon and jet production. The method consists in parametrizing
the parton densities at some input scale $Q_0$ and then adjusting the 
parameters to fit the data. The parameters are usually constrained by imposing
the positivity of the parton densities $(f_{a}(x, \mu^2) \geq 0)$ and
the momentum sum rule $(\sum_a \int_0^1 dx \,x \,f_{a}(x, \mu^2) =1)$.

\begin{figure}
  \centerline{
    \setlength{\unitlength}{1cm}
    \begin{picture}(0,6)
       \put(0,0){\includegraphics{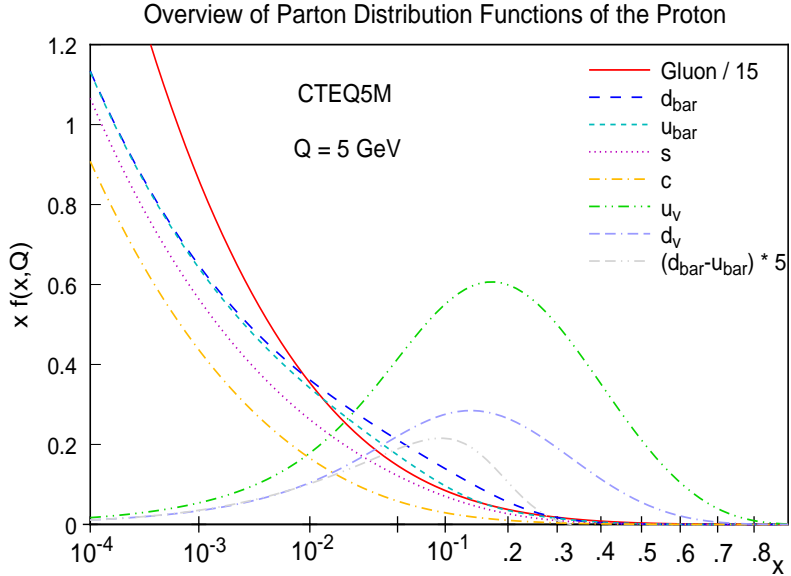}}
    \end{picture}}
\vspace*{1.3cm}    
\caption{Typical $x$-shape of the parton densities (set CTEQ5M at $Q=5$~GeV).
\label{pdfplot}}
\end{figure}

The present knowledge on the parton densities of the proton is reviewed in
Refs.~[\ref{proclhc}, \ref{lhcpdf}].
Their typical behaviour is shown in 
Fig.~\ref{pdfplot}. 
All densities decrease at large $x$. At small $x$ the valence quark densities
vanish and the gluon density dominates. The sea-quark densities also increase
at small $x$ because they are driven by the strong rise of the gluon density
and the splitting of gluons in $q {\bar q}$ pairs. Note that the quark
densities are not flavour-symmetric either in the valence sector
$(u_v \neq d_v)$ or in the sea sector $({\bar u} \neq {\bar d})$. 

In addition to having the best estimate of the parton densities, it is 
important to quantify the corresponding uncertainty. This is a difficult issue.
The uncertainty depends on the kinematic range in $x$ and $Q^2$. Moreover,
it cannot be reliably estimated by simply comparing the parton densities
obtained by different global fits. In fact, a lot of common systematic and
common assumptions affect the global-fit procedures.
Recent attempts to
obtain parton densities with error bands that take into account correlations
between experimental errors are described in Refs.~[\ref{proclhc}, 
\ref{pdfrball}].
Some important theoretical uncertainties that are still to be understood
are also discussed in Ref.~[\ref{proclhc}]. 

The overall conclusion is that
the quark densities\footnote{Uncertainties on the determination of the quark
densities at very high $x$ are discussed in Refs.~[\ref{bodek}, 
\ref{Melnitchouk}, \ref{highxcteq}].}
are reasonably well constrained and determined by
DIS and DY processes, while the gluon density is certainly more 
uncertain [\ref{pdffit}, \ref{gluoncteq}].
At small $x$ $(x \ltap 10^{-3})$, the gluon density $f_g$ is at present 
constrained by a {\em single} process, namely DIS at HERA. Thus, 
large higher-order corrections of the type $(\as \ln 1/x)^n$ could possibly 
affect the extraction of $f_g$.
{\em Assuming}
that $f_g$ is well determined at small $x$, the momentum sum rule reasonably
constrains $f_g$ at intermediate values of $x$ $(x \sim 10^{-2})$.
Jet production at the Tevatron at low to moderate values of the
jet transverse energy $E_T$ can also be useful
in constraining the gluon distribution in the range $0.05 \ltap x \ltap 0.2$.
At large $x$ $(x \sim 10^{-1})$, the most sensitive process to $f_g$
is prompt-photon production. Since, at present, prompt-photon data are not well
described/predicted by perturbative QCD calculations, they cannot be used for a
precise determination of $f_g$. Further discussion on these points is given
in Sect.~\ref{secgdensity}.
 
\begin{figure}
  \centerline{
    \setlength{\unitlength}{1cm}
    \begin{picture}(0,6)
       \put(0,0){\includegraphics{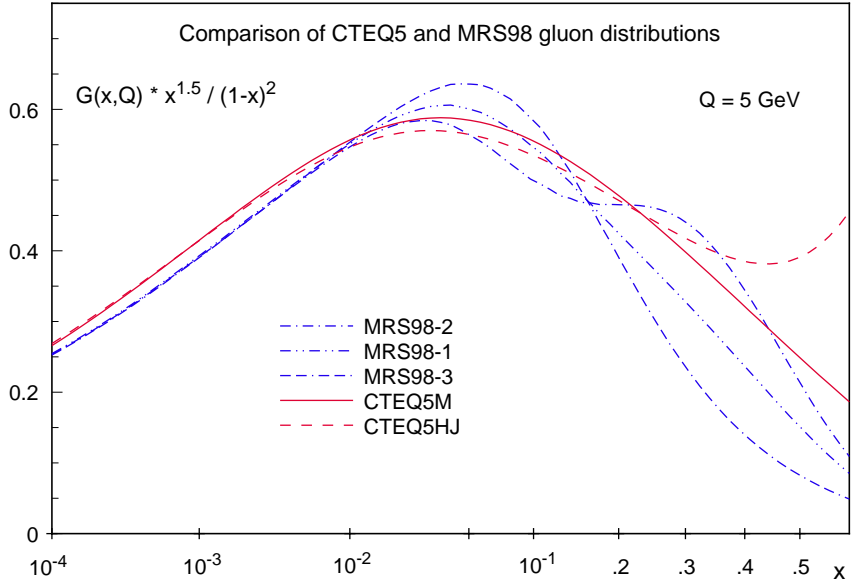}}
    \end{picture}}
\vspace*{1.5cm}    
\caption{Comparison between the gluon densities of the CTEQ and MRST groups.
\label{differentfg}}
\end{figure}

The conclusion that the gluon density is not well known can also be drawn
by inspection (see Fig.~\ref{differentfg})
of the differences between the most updated analyses performed
by the CTEQ Collaboration and the MRST group.

The differences between the MRST gluons and the CTEQ ones are due to the fact
that the two groups used different data sets.
The various gluon densities are very
similar at small $x$, because in this region both groups used the HERA data.
The MRST group includes prompt-photon data in the global fit: these data 
constrain the gluon directly at $x \gtap 10^{-1}$ and indirectly (by the momentum
sum rule) at $x \sim 10^{-2}$. The CTEQ group does not use prompt-photon data,
but it includes Tevatron data on the one-jet inclusive cross section. 
These data
give a good constraint on $f_g$ in the region $10^{-2} < x < 10^{-1}$.
 
There are also differences within the MRST and CTEQ sets. The various gluon
densities of the MRST set correspond to different values of the non-perturbative
transverse-momentum smearing that can be introduced to describe the differences
among the prompt-photon data that are available at several 
centre-of-mass energies. The CTEQ5M and CTEQ5HJ gluons correspond to different
assumptions on the parametrization of the functional form of $f_g(x,Q_0^2)$ at
large $x$; the CTEQ5M set corresponds to the minimum-$\chi^2$ solution of 
the fit while the CTEQ5HJ set (with a slightly higher $\chi^2$) provides 
the best fit to the high-$E_T$ tail of the CDF {\em and} D0 jet cross sections. 

\begin{figure}
  \centerline{
    \setlength{\unitlength}{1cm}
    \begin{picture}(0,6)
       \put(0,0){\includegraphics{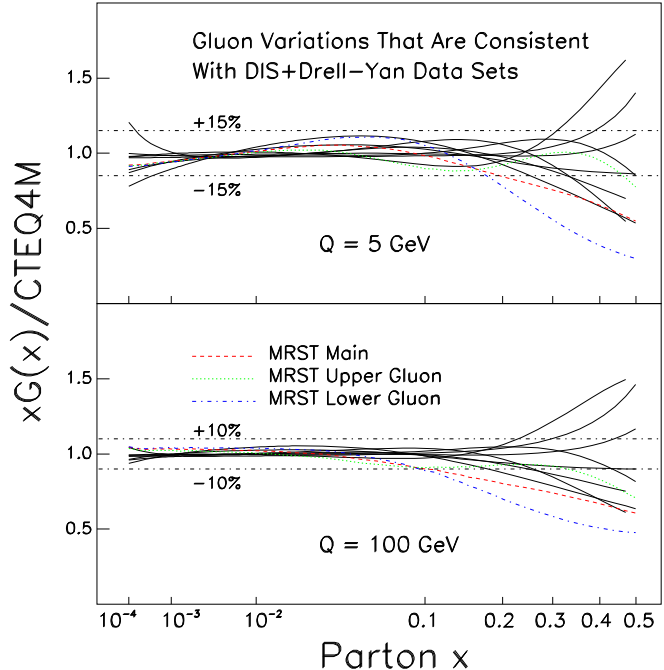}}
    \end{picture}}
\vspace*{2.8cm}    
\caption{A picture of the gluon density uncertainty. The continuous (black)
lines refer to gluon densities that are constrained only by DIS and DY data.
The dashed (coloured) lines refer to gluon densities of the MRST group, which
uses also prompt-photon data.
\label{fguncertainty}}
\end{figure}

This brief illustration shows that the differences in the most recent
parton densities are mainly due to either inconsistencies between data sets
and/or poor theoretical understanding of them.
A more quantitative picture of the dependence on $x$ and $Q^2$
of the gluon density uncertainty is presented in Fig.~\ref{fguncertainty}.

We can see that the DIS and DY data sets weakly constrain $f_g$ for
$x \gtap 10^{-1}$. Since the AP splitting functions lead to negative scaling
violation at large $x$, when $f_g(x,Q^2)$ is evolved at larger scales $Q$
according to Eq.~(\ref{evequa}) the gluon uncertainty 
is diluted: it propagates at smaller values of $x$ and its size 
is reduced at fixed $x$.

\begin{figure}
  \centerline{
    \setlength{\unitlength}{1cm}
    \begin{picture}(0,7)
       \put(0,0){\includegraphics{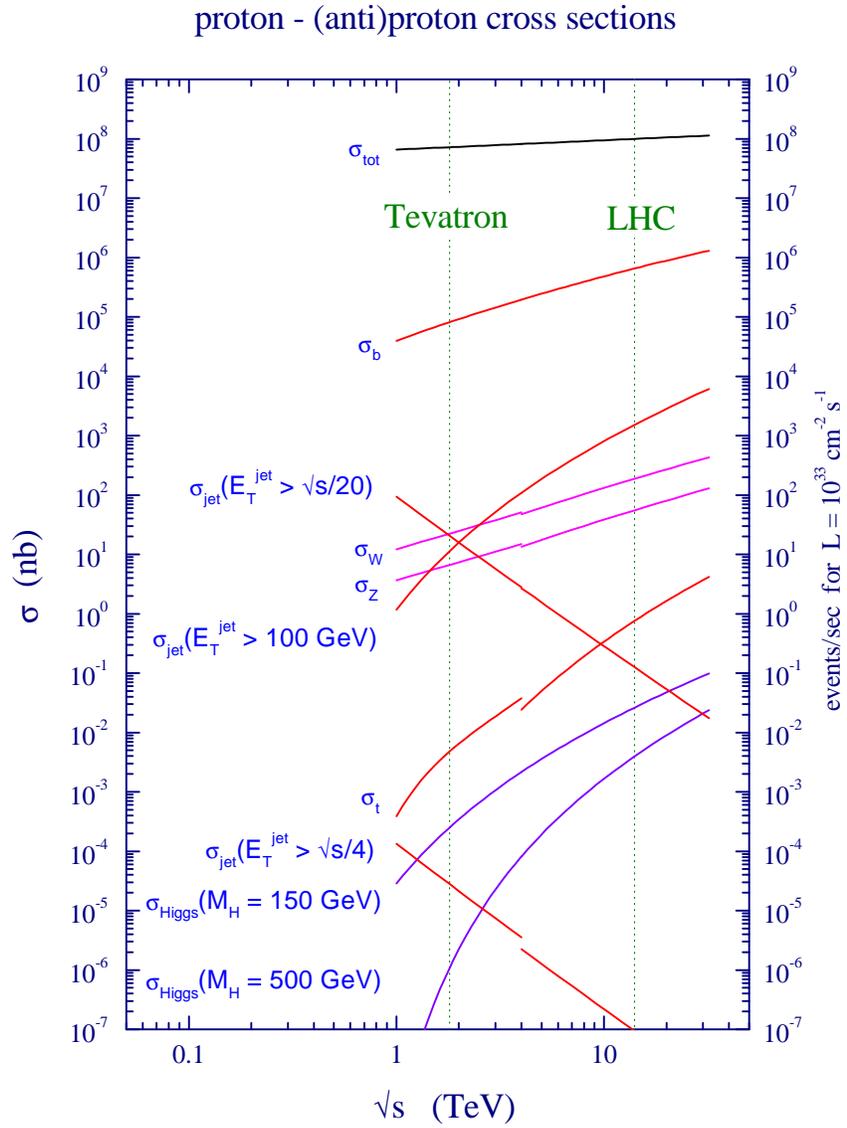}}
    \end{picture}}
\vspace*{4.7cm}    
\caption{QCD predictions for hard-scattering cross sections at the Tevatron 
and the LHC.
\label{figxs}}
\end{figure}

Figure~\ref{figxs}
shows the typical predictions for hard-scattering cross sections at the 
Tevatron and the LHC, as obtained by using the parton densities of the MRST 
set. These predictions have to be supplemented with the corresponding
uncertainties [\ref{proclhc}]
coming from the determination of the parton 
densities and from perturbative corrections beyond the NLO.
Owing to the
increased centre-of-mass energy and to QCD scaling violation 
(see Fig.~\ref{fguncertainty}), the kinematic region with small uncertainties
is larger at the LHC than at the Tevatron. 

For most of the QCD processes at the LHC, the uncertainty from
the parton densities is smaller than $\pm 10\%$ and, in particular,
it is smaller than the uncertainty from higher-order corrections. Some 
relevant exceptions are the single-jet, $W/Z$ and top quark cross sections.
In the case of the single-jet inclusive cross section at high $E_T$
$(E_T \gtap 2$~TeV), the uncertainty from the poorly known gluon density at 
high $x$ is larger than that $(\sim \pm 10\%)$ from higher-order corrections.
The $W$ and $Z$ production cross sections are dominated by $q{\bar q}$
annihilation. Since the quark densities are well known, the ensuing uncertainty
on the $W/Z$ cross section is small $(\sim \pm 5\%)$. Nonetheless, in this case
the uncertainty from higher-order corrections is even smaller, since the
partonic cross sections for the DY process are known [\ref{NNLODY}] at the
next-to-next-to-leading order (NNLO) in perturbation theory. In the case
of top-quark production at the LHC, the gluon channel dominates and leads to
an uncertainty of $\pm 10\%$ on the total cross section. Also for this process,
however, the perturbative component is known beyond the NLO. Including
all-order resummation of soft-gluon contributions [\ref{txs}], 
the estimated uncertainty from unknown higher-order corrections is 
approximately $\pm 5\%$ [\ref{txs}, \ref{proclhc}].

\section{The gluon density issue}
\label{secgdensity}

At present, the processes\footnote{The r\^ole of jet production at the Tevatron
has briefly been recalled in Sect.~\ref{secpdf}, and it is discussed in detail
in Ref.~[\ref{lhcpdf}].} that are, in principle,
most sensitive to the gluon density are
DIS at HERA, $b$-quark production at the Tevatron, and prompt-photon production
at fixed-target experiments. These processes constrain $f_g$
for $x \ltap 10^{-3}$, $x \sim 10^{-3}$--$10^{-2}$ and $x \gtap 10^{-1}$,
respectively.
Nonetheless, the gluon density is, in practice, not well determined. 
The issue (or, perhaps, the puzzle)
is that from a phenomenological viewpoint the standard theory, namely
perturbative QCD at NLO, works pretty well for $x \ltap 10^{-3}$ but not
so well at larger values of $x$, while from theoretical arguments we should
expect just the opposite to happen. This issue is discussed below mainly in its
perturbative aspects. We should however keep it in mind that all these processes
are dominated by hard-scattering scales $Q$ of the order of few GeV. Different
types of non-perturbative contributions can thus be important.

From the study of DIS at HERA we can extract information on the gluon and
sea-quark densities of the proton. The main steps in the QCD analysis of the
structure functions at small values of the Bjorken variable $x$
are the following.
The measurement of the proton structure function
$F_2(x,Q^2) \sim q_S(x,Q^2)$ directly determines the sea-quark density
$q_S= x (f_q + f_{\bar q})$. Then, the DGLAP evolution equation (\ref{evequa})
or, more precisely, the following equations
(the symbol $\otimes$ denotes the convolution integral with respect to $x$):
\beeq
\label{df2}
dF_2(x,Q^2) / d\ln Q^2 
&\sim& P_{qq} \otimes q_S +  P_{qg} \otimes g \;\;, \\
\label{dfg}
d g(x,Q^2)/ d\ln Q^2 
&\sim& P_{gq} \otimes q_S +  P_{gg} \otimes g \;\;, 
\eeeq
are used to extract a gluon density $g(x,Q^2)= x f_g(x,Q^2)$ that agrees with 
the measured scaling violation in $dF_2(x,Q^2) / d\ln Q^2$ 
(according to Eq.~(\ref{df2}))
and fulfils the self-consistency equation (\ref{dfg}). 

The perturbative-QCD 
ingredients in this analysis are the AP splitting functions 
$P_{ab}(\as,x)$.
Once they are known (and only then), the non-perturbative gluon density
can be determined. 

The standard perturbative-QCD framework to extract $g(x,Q^2)$ consists 
in using the truncation of the AP splitting functions at the NLO. This
approach has been extensively compared with structure function
data over the last few years and it gives a good description of the HERA data,
down to low values of $Q^2 \sim 2 \, {\rm GeV^2}$. The NLO QCD fits simply 
require a slightly steep input gluon density at these low momentum scales.
Typically [\ref{pdffit}], we have $g(x,Q_0^2) \sim x^{-\lambda}$, 
with $\lambda \sim 0.2$ at $Q_0^2 \sim 2 \, {\rm GeV^2}$, and the data
constrain $g(x,Q_0^2)$ with an uncertainty of approximately $\pm 20\%$.

Although it is phenomenologically successful, the NLO approach is
not fully satisfactory from a theoretical viewpoint.
The truncation of the splitting functions
at a fixed perturbative order is equivalent to assuming that the dominant
dynamical mechanism leading to scaling violations is the evolution of
parton cascades with strongly-ordered transverse momenta. However, at high 
energy this evolution takes place over large rapidity intervals $(\Delta  y \sim
\ln1/x)$ and diffusion in transverse momentum becomes relevant. Formally, this
implies that higher-order corrections to $P_{ab}(\as,x)$ are logarithmically
enhanced:
\beq
\label{plog}
P_{ab}(\as,x) \sim \frac{\as}{x} + \frac{\as}{x} \;( \as \ln x ) + \dots
+ \frac{\as}{x} \;( \as \ln x )^n + \dots \;\;.
\eeq
At asymptotically small values of $x$, resummation of these corrections is 
mandatory to obtain reliable predictions.

Small-$x$ resummation is, in general, accomplished by the BFKL 
equation~[\ref{BFKL}]. In the context of structure-function
calculations, the BFKL equation provides us with improved expressions
of the AP splitting functions $P_{ab}(\as,x)$, in which the leading logarithmic
(LL) terms $(\as \ln x )^n$, the next-to-leading logarithmic (NLL) terms 
$\as (\as \ln x )^n$, and so forth, are systematically summed to all orders $n$
in $\as$. The present theoretical status of small-$x$ resummation is discussed
in Ref.~[\ref{proclhc}].
Since in the small-$x$ region the gluon channel 
dominates, only the gluon splitting functions $P_{gg}$ and $P_{gq}$ contain
LL contributions. These are known [\ref{BFKL}, \ref{Jar}] to be positive
but numerically smaller than naively expected
(the approach to the asymptotic regime is
much delayed by cancellations of logarithmic
corrections that occur at the first perturbative orders in $P_{gg}$ and 
$P_{gq}$). The NLL terms in the quark splitting functions $P_{qg}$ and 
$P_{qq}$ are known [\ref{CH}] and turn out to be positive and large.
 A very important progress is the recent calculation
[\ref{fadin}, \ref{CC}] of the NLL terms in $P_{gg}$, which are found to
be negative and large. The complete NLL terms in $P_{gq}$ are still unknown.

The results of Refs.~[\ref{fadin}, \ref{CC}], the large size of the NLL terms
and the alternating sign (from the LL to the NLL order and from
the gluon to the quark channel) of the resummed
small-$x$ contributions have 
prompted a lot of activity (see the list of references in Ref.~[\ref{proclhc}]) 
on the conceptual basis and the phenomenological implications of small-$x$ 
resummation. This activity is still in progress and definite quantitative
conclusions on the impact of small-$x$ resummation at HERA cannot be drawn yet.

\begin{figure}
  \centerline{
    \setlength{\unitlength}{1cm}
    \begin{picture}(0,6)
       \put(0,0){\includegraphics{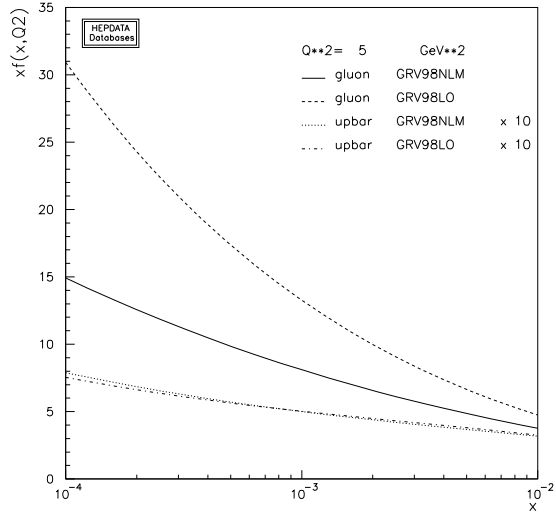}}
    \end{picture}}
\vspace*{0.7cm}    
\caption{Comparison between the LO (GRV98LO) and NLO (GRV98NLM) GRV 
parametrizations of the gluon and sea-quark densities at $Q^2=5$~GeV$^2$.
\label{figgrv}}
\end{figure}
 
At the same time, the capability of the fixed-order approach to produce a good 
description of the proton
structure function $F_2(x,Q^2)$ at HERA cannot be used to conclude
that the small-$x$ behaviour of the gluon density is certainly well determined.
In fact, by comparing LO and NLO results, we could argue that the ensuing
theoretical uncertainty on $f_g$ is sizeable [\ref{lp97}].
Going from LO to NLO, we can obtain stable predictions for $F_2$,
but we have to vary the gluon density a lot.
As shown in Fig.~\ref{figgrv}, the NLO gluon density sizeably differs from its
LO parametrization, not only in absolute normalization but also in $x$-shape.
For instance, at $x=10^{-4}$ and $Q^2=5~{\rm GeV}^2$ the NLO gluon is a factor
of 2 smaller than the LO gluon.
This can be understood~[\ref{sdis}] from the fact that the scaling violation of 
$F_2$ is produced by the convolution $P_{qg} \otimes g$ (see the right-hand 
side of Eq.~(\ref{df2})). The quark splitting function $P_{qg}$ behaves as
\beq
\label{pqg}
P_{qg}(\as,x) \simeq \as P_{qg}^{(LO)}(x) \left[ 1 + 2.2 \frac{C_A \as}{\pi}
\frac{1}{x} + \dots \right] \;\;,
\eeq
where the LO term $P_{qg}^{(LO)}(x)$ is flat at small $x$, whereas the NLO
correction is steep. To obtain a stable evolution of $F_2$, the NLO steepness 
of $P_{qg}$ has to be compensated by a gluon density that is less steep 
at NLO than at LO. This has to be kept in mind when
concluding on the importance of small-$x$ resummation because
the NLO steepness of $P_{qg}$ is the lowest-order manifestation of BFKL 
dynamics in the quark channel. 

In the large-$x$ region, there is a well-known correlation between
$\as$ and $f_g$. At small $x$, there is an analogous strong correlation
between the $x$-shapes of $P_{qg}$ and $f_g$. In the fixed-order QCD analysis
of $F_2$, large NLO perturbative corrections at small $x$ can be balanced by
the extreme flexibility of parton density parametrizations. It is difficult
to disentangle this correlation between process-dependent
perturbative contributions and non-perturbative parton densities from the study 
of a single quantity, as in the case of $F_2$ at HERA. The uncertainty on the
gluon density at small $x$, as estimated from the NLO QCD fits of the
HERA data, is evidently only a lower limit on the actual uncertainty on $f_g$.

\begin{figure}
  \centerline{
    \setlength{\unitlength}{1cm}
    \begin{picture}(0,6)
       \put(0,0){\includegraphics{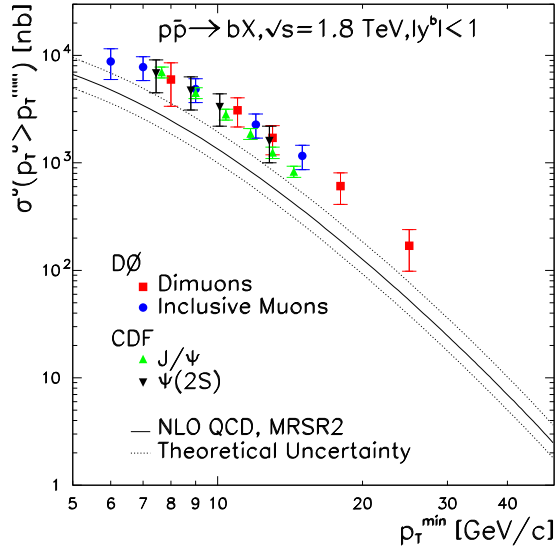}}
    \end{picture}}
\vspace*{1.7cm}    
\caption{Comparison between Tevatron data and NLO QCD for $b$-quark production
[\ref{baarmand}]. The band is obtained by varying factorization
and renormalization scales in the NLO calculation.
\label{figbtev}}
\end{figure}

The production of $b$ quarks at the Tevatron is also sensitive
to the gluon density at relatively small values of $x$. The comparison
between Tevatron data and perturbative-QCD predictions at NLO [\ref{hqnlo}] 
is shown in Fig.~\ref{figbtev}. 
Using standard sets of parton densities, the theoretical predictions 
typically underestimate the measured cross section by a factor of 2.
This certainly is disappointing, although justifiable by the large theoretical
uncertainty of the perturbative calculation [\ref{bquarkrev}]. A lower limit
on this uncertainty can be estimated by studying the scale dependence and the
convergence of the perturbative expansion.
Varying the factorization
and renormalization scales by a factor of four around the $b$-quark mass $m_b$,
the NLO cross section varies by a factor of almost 2 at the Tevatron
and by a factor of 4--5 at the LHC [\ref{proclhc}]. 
Similar factors are obtained by considering
the ratio of the NLO and LO cross sections.

The present theoretical predictions for $b$-quark production at hadron 
colliders
certainly need to be improved [\ref{proclhc}]. Since the hard scale 
$Q \sim m_b$ is not very large, a possible improvement regards estimates of 
non-perturbative contributions (for instance, effects of the fragmentation
of the $b$-quark and of the intrinsic transverse momentum of the colliding 
partons). As for the evaluation of perturbative contributions at higher 
orders, 
the resummation of logarithmic terms of the type $\as^n \ln^n(p_t/m_b)$ is 
important [\ref{ptbres}] when the transverse momentum $p_t$ of the $b$ quark 
is much larger than $m_b$. The resummation of small-$x$ logarithmic 
contributions $\as^n \ln^n x$ can also be relevant, because 
$x \sim 2m_b/{\sqrt S}$ is as small as $\sim 10^{-3}$ at the Tevatron and
as $\sim 10^{-4}$ at the LHC. The theoretical tool to perform this resummation,
namely the $k_{\perp}$-factorization approach [\ref{ktfac}], is available.
Updated phenomenological studies based on this tool and on the information
from small-$x$ DIS at HERA would be interesting.

\begin{figure}
  \centerline{
    \setlength{\unitlength}{1cm}
    \begin{picture}(0,6)
       \put(0,0){\includegraphics{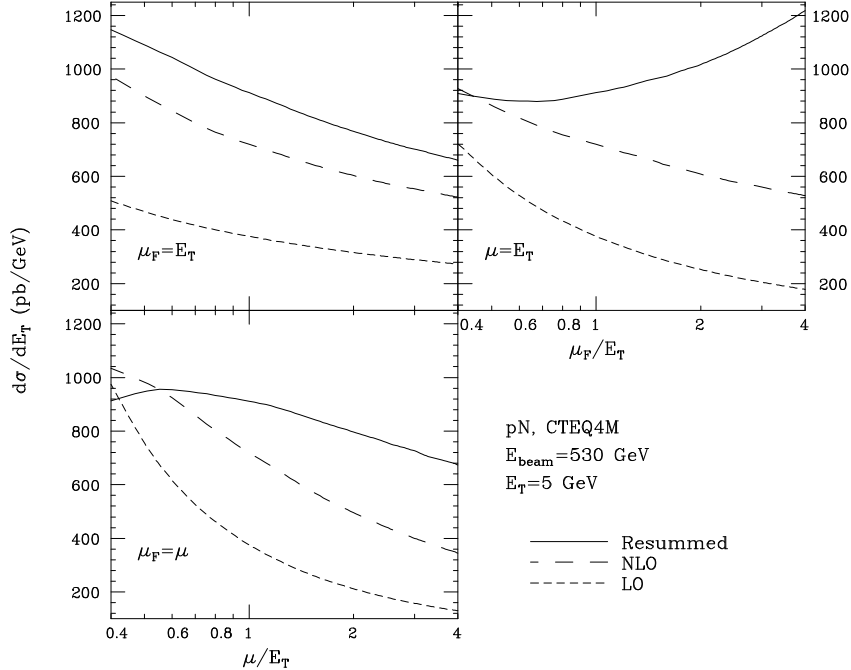}}
    \end{picture}}
\vspace*{2.5cm}    
\caption{The dependence on the factorization $(\mu_F)$ and renormalization
$(\mu_R=\mu)$ scale of the LO and NLO prompt-photon cross section 
$d\sigma/dE_T$ in $pN$ collisions at $E_T=5$~GeV and ${\sqrt S}=31.6$~GeV. 
The resummed calculation is discussed in Sect.~\ref{secsoftg}.
\label{figgamma5}}
\end{figure}

Prompt-photon production at fixed-target experiments is sensitive to the
behaviour of the gluon density at large $x$ $(x \gtap 0.1)$. The theoretical
predictions for this process, however, are not very accurate. 
Figure~\ref{figgamma5} shows the 
factorization- and renormalization-scale dependence of the perturbative cross
section for the case of the E706 kinematics. If the scale is varied by a factor
of 4 around the transverse energy $E_T$ of the prompt photon, the LO
cross section varies by a factor of almost 4. Going to NLO 
[\ref{promptgnlo}] the situation
improves, but not very much, because the NLO cross section still varies by 
a factor of about 2.

\begin{figure}
  \centerline{
    \setlength{\unitlength}{1cm}
    \begin{picture}(0,6)
       \put(0,0){\includegraphics{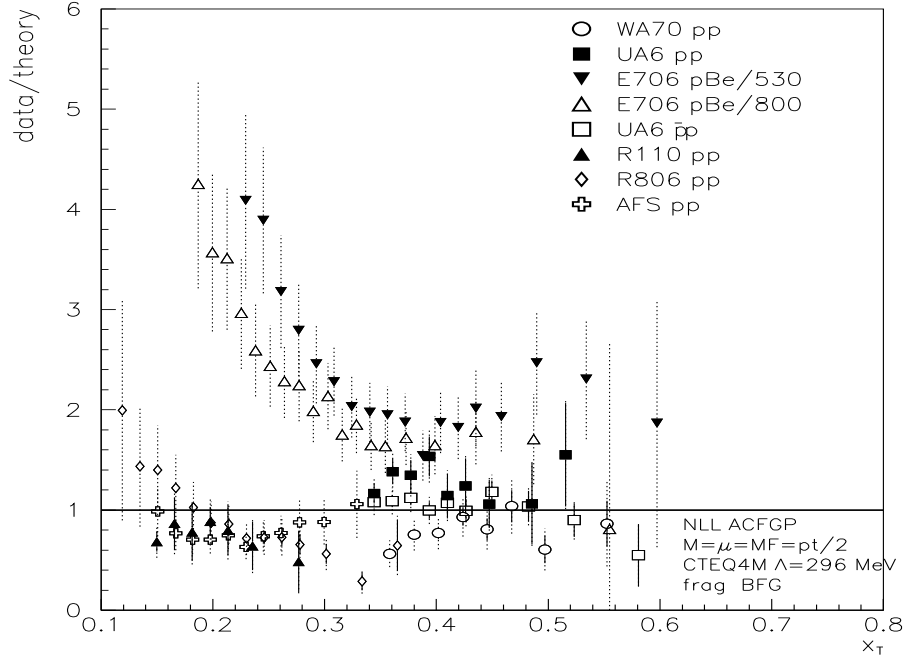}}
    \end{picture}}
\vspace*{2.5cm}    
\caption{A comparison between NLO QCD calculations and data from the ISR and 
fixed-target experiments for the prompt-photon distribution $d\sigma/dE_T$ 
$(x_T=2E_T/{\sqrt S})$.
\label{figannecy}}
\end{figure}

A detailed comparison between NLO QCD calculations and data from the ISR and
fixed-target experiments has recently been performed in
Ref.~[\ref{annecygamma}]. As shown in Fig.~\ref{figannecy}, the overall
agreement with the theory is not satisfactory, even taking into account the
uncertainty coming from scale variations in the theoretical predictions.
Modifications of the gluon density can improve the agreement with some data
sets only at the expense of having larger disagreement with other data sets.
The differences between experiments at similar centre-of-mass energies
(see, for instance, E706 pBe/530 at ${\sqrt S}=31.6~{\rm GeV}$ and
WA70 pp at ${\sqrt S}=23~{\rm GeV})$ are much larger than expected from 
perturbative scaling violations. This can possibly suggest [\ref{annecygamma}]
inconsistencies of experimental origin. 

Another (not necessarily alternative)
origin of the differences between data and theory could be the presence of
non-perturbative effects that are not included in the NLO perturbative
calculation. This explanation has been put forward in 
Refs.~[\ref{E706}, \ref{ktcteq}] by introducing some amount of 
intrinsic\footnote{To be precise, in Ref.~[\ref{ktcteq}]
the $\langle k_{\perp} \rangle$ of the colliding partons is not called 
`intrinsic', but it is more
generically called the $\langle k_{\perp} \rangle$ `from initial-state
soft-gluon radiation'.} 
transverse momentum $\langle k_{\perp} \rangle$
of the colliding partons. Owing to the steeply
falling $E_T$ distribution $(d\sigma/dE_T \propto 1/E_T^7)$ of the 
prompt photon, even a small transverse-momentum 
kick\footnote{The $E_T$ distribution of the single-photon is not calculable
down to $E_T=0$ or, in other words, $d\sigma/dE_T$ is not integrable in the
entire kinematic range of $E_T$. Thus, the intrinsic $\langle k_{\perp} \rangle$
of the incoming partons does not simply produce a shift of events from
the low-$E_T$ to the high-$E_T$ region. For this reason, the terminology 
`$\langle k_{\perp} \rangle$ kick' seems to be more appropriate than
`$\langle k_{\perp} \rangle$ smearing'.}
can indeed produce
a large effect on the cross section, in particular, 
at small values of $E_T$. Phenomenological investigations [\ref{ktcteq}]
show that this additional $\langle k_{\perp} \rangle$ kick can lead to a
better agreement between calculations and data. The E706 data suggest the
value $\langle k_{\perp} \rangle \sim 1.2~{\rm GeV}$, the WA70 data prefer
no $\langle k_{\perp} \rangle$, and the UA6 data in the intermediate range of
centre-of-mass energy $({\sqrt S}=24.3~{\rm GeV}$) may prefer an intermediate
value of $\langle k_{\perp} \rangle$. Similar conclusions are obtained in the 
analysis by the MRST group [\ref{pdffit}].

A precise physical understanding of $\langle k_{\perp} \rangle$ 
effects is still missing. On one side,
since the amount of $\langle k_{\perp} \rangle$ 
suggested by prompt-photon data varies with ${\sqrt S}$, it is difficult to
argue that the transverse momentum is really `intrinsic' and has an
entirely non-perturbative origin. On the other side, in the case of the
inclusive production of a single photon, 
a similar effect cannot
be justified by higher-order {\em logarithmic}
corrections produced by perturbative
soft-gluon radiation (see Sect.~\ref{secsoftg}). A lot of model-dependent
assumptions (and ensuing uncertainties) certainly enter in the present 
implementations of the $\langle k_{\perp} \rangle$ kick. 
A general framework to {\em consistently} include non-perturbative 
transverse-momentum effects in perturbative calculations is not yet available. 
Recent proposals with this aim are presented in Refs.~[\ref{ktmartin}]
and [\ref{ktsterman}].

Further studies on the consistency between different
prompt-photon experiments and on the issue of
intrinsic-$\langle k_{\perp} \rangle$ effects in hadron--hadron collisions 
are necessary. Owing to the present theoretical
(and, possibly, experimental) uncertainties, it is difficult to use
prompt-photon data to accurately determine the gluon density at large $x$. 
Other recent theoretical improvements, such as soft-gluon resummation,
of the perturbative calculations for prompt-photon production at large 
$x_T=2E_T/{\sqrt S}$ are discussed in Sect.~\ref{secsoftg}. 

Studies of other single-particle inclusive cross sections, 
such as $\pi^0$ cross sections [\ref{ktcteq}, \ref{annecypi}, \ref{jh}], 
can be valuable to
constrain the parton densities and could possibly help to clarify some of
the experimental and theoretical issues arisen by prompt-photon production.
 
\setcounter{footnote}{0}
\section{Partonic cross sections: fixed-order expansions, \\
resummed calculations,
parton showers}
\label{secpxs}

The calculation of hard-scattering cross sections according
to the factorization formula (\ref{factfor}) requires the knowledge of the 
partonic cross sections ${\hat \sigma}$, besides that of the parton densities.
The partonic cross sections are 
usually computed by truncating their perturbative expansion at a fixed order 
in $\as$:
\beeq
 \label{pertex}
\!\!\!\!\!\!
{\hat \sigma}(p_1,p_2;Q, \{Q_1, \dots \}; \mu_F^2) \!\!\!\!&=&\!\!\!\!
\as^k(\mu_R^2) \left\{
{\hat \sigma}^{(LO)}(p_1,p_2;Q, \{Q_1, \dots \}) \right.  \\
&~& \;\;\;\;\;\; \;\;\;\; + \,\as(\mu_R^2) \;
{\hat \sigma}^{(NLO)}(p_1,p_2;Q, \{Q_1, \dots \}; \mu_R^2; \mu_F^2) \nn \\
&~& \;\;\;\;\;\; \;\;\;\; +\left. \! \as^2(\mu_R^2) \;
{\hat \sigma}^{(NNLO)}(p_1,p_2;Q, \{Q_1, \dots \}; \mu_R^2; \mu_F^2) +
\dots \right\} \,.\nn
\eeeq
The scale $\mu_R$ is the arbitrary renormalization scale introduced to define
the perturbative expansion. Although the `exact' partonic cross section on 
the left-hand side of Eq.~(\ref{pertex}) does not depend on $\mu_R$, each term
on the right-hand side (and, hence, any fixed-order truncation) separately
depends on it.

The LO (or tree-level) term ${\hat \sigma}^{(LO)}$ gives only an estimate
of the order of magnitude of the partonic cross section, because
at this order $\as$ is not unambiguously defined. Equivalently, we can say
that since ${\hat \sigma}^{(LO)}$ does not 
depend on $\mu_R$, the size of its contribution can be varied quite 
arbitrarily by changing $\mu_R$ in its coefficient $\as^k(\mu_R^2)$. The strong
coupling $\as$ can be precisely defined only starting from NLO. A `reliable'
estimate of the central value of
${\hat \sigma}$ thus requires the knowledge of (at least)
the NLO term
${\hat \sigma}^{(NLO)}$. This term explicitly depends on $\mu_R$ and this
dependence begins to compensate that of $\as(\mu_R^2)$. 

In general, the
$n$-th term in the curly bracket of Eq.~(\ref{pertex}) contains contributions
of the type $( \as(\mu_R^2) \ln Q/\mu_R )^n$.
If $\mu_R$ is very different
from the hard scale $Q$, these contributions become large and spoil the 
reliability of the truncated expansion (\ref{pertex}).  
Thus, in practical applications the scale $\mu_R$ should be set approximately 
equal to the hard scale $Q$. As mentioned in Sect.~\ref{secgdensity}, 
variations of $\mu_R$ around this central value
are typically used to set a lower limit on the theoretical uncertainty
of the perturbative calculation.

A better estimate of the accuracy of any perturbative expansion is obtained
by considering the effect of removing the last perturbative term that has been
computed. Since $\as$ can be precisely defined only at NLO, this 
procedure can consistently be applied to Eq.~(\ref{pertex}) only as
from its NNLO term. A `reliable' estimate of the theoretical error on
${\hat \sigma}$ thus requires the knowledge of 
the NNLO term ${\hat \sigma}^{(NNLO)}$ in Eq.~(\ref{pertex}).

The LO and NLO approximations of ${\hat \sigma}$ are used at present
in (most of) the fixed-order QCD calculations. Prospects towards NNLO
calculations of partonic cross sections and AP splitting functions
are reviewed in Refs.~[\ref{proclhc}, \ref{delduca}].

The fixed-order expansion (\ref{pertex}) provides us with a well-defined and
systematic framework to compute the partonic cross section 
${\hat \sigma}(p_1,p_2;Q, \{Q_1, \dots \}; \mu_F^2)$ of any
hard-scattering process that is sufficiently inclusive or, more precisely,
that is defined in an infrared- and collinear-safe manner. However,
the fixed-order expansion is reliable only when all the kinematical scales
$Q, \{Q_1, \dots \}$ are of the same order of magnitude. 
When the hard-scattering process involves two (or several) very different 
scales, say $Q_1 \gg Q$, the ${\rm N}^n{\rm LO}$ term in Eq.~(\ref{pertex}) can 
contain double- and single-logarithmic contributions of the type
$(\as L^2)^n$ and $(\as L)^n$ with $L=\ln (Q_1/Q) \gg 1$. These terms spoil
the reliability of the fixed-order expansion and have to be summed to all orders
by systematically defining logarithmic expansions (resummed calculations).

Typical large logarithms, $L=\ln Q/Q_0$, are those related to the evolution of 
the parton densities from a low input scale $Q_0$ to the hard-scattering scale
$Q$. These logarithms are produced by collinear radiation from the colliding 
partons and give single-logarithmic contributions. They never explicitly appear
in the calculation of the partonic cross section, because they are 
systematically (LO, NLO and so forth) resummed in the evolved parton densities 
$f(x,Q^2)$ by using the DGLAP equation (\ref{evequa}).

Different large logarithms, $L=\ln Q/{\sqrt S}$, appear when the centre-of-mass
energy ${\sqrt S}$ of the collision is much larger than the hard scale $Q$. 
These small-$x$ $(x=Q/{\sqrt S})$ logarithms are produced by multiple radiation
over the wide rapidity range that is available at large energy. They usually
give single-logarithmic contributions that can be resummed by using the BFKL
equation. BFKL resummation is relevant
to DIS structure functions at small values of the Bjorken variable $x$ 
(see Sect.~\ref{secgdensity}) and it can also be important at the LHC for the 
production of $b$ quarks and of prompt photons at relatively low $E_T$.

Another class of large logarithms is associated to the bremsstrahlung
spectrum of soft gluons. Since soft gluons can be radiated collinearly,
they give rise to double-logarithmic contributions to the partonic cross
section:
\beq
\label{logxs}
{\hat \sigma} \sim \as^k {\hat \sigma}^{(LO)} 
\left\{ 1 + \sum_{n=1}^{\infty} \as^n \left( C_{2n}^{(n)} L^{2n} +  
C_{2n-1}^{(n)} L^{2n-1}
+ C_{2n-2}^{(n)} L^{2n-2} + \dots \right) \right\} \;\;.
\eeq
Soft-gluon resummation is discussed in Sect.~\ref{secsoftg}.

A related approach to evaluate higher-order contributions to the 
partonic cross sections is based on Monte Carlo parton showers 
(see [\ref{book}] and the updated list of references in 
[\ref{proctev}, \ref{proclhc}]). 
Rather than computing exactly ${\hat \sigma}^{(NLO)}$,
${\hat \sigma}^{(NNLO)}$ and so forth,
the parton shower gives an all-order approximation
of the partonic cross section in the soft and collinear regions.
In this respect, the computation of the partonic cross sections performed by
parton showers is somehow similar to that obtained by soft-gluon resummed 
calculations. There is, however, an important conceptual difference between
the two approaches. This difference and the limits of applicability of
the parton-shower method are briefly recalled below. Apart from these limits,
parton-shower calculations can give some advantages. Multiparton kinematics
can be treated exactly. The parton shower can be 
supplemented with models of non-perturbative effects
(hadronization, intrinsic $k_{\perp}$, soft underlying event)
to provide a complete description of the hard-scattering 
process at the hadron level. 

For a given cross section, resummed calculations can in 
principle be performed to any logarithmic accuracy. The logarithmic accuracy 
achievable by parton showers is instead intrinsically limited by quantum 
mechanics. The parton-shower algorithms are probabilistic.
Starting from the LO cross 
section, the parton shower generates multiparton final states according to
a probability distribution that approximates the {\em square} of the QCD 
matrix elements.
The approximation is based on 
the universal (process-independent) factorization properties of multiparton 
matrix elements in the soft and collinear limits. Although the matrix element
does factorize, its square contains quantum interferences, which are not 
positive-definite and, in general, cannot be used to define probability 
distributions. To leading infrared accuracy,
this problem is overcome by exploiting QCD coherence (see
Refs.~[\ref{book}, \ref{Dokbook}, \ref{BCM}] and referencees therein):
soft gluons radiated at
large angle from the partons involved in the LO subprocess destructively
interfere. This quantum mechanical effect can be simply implemented
by enforcing an angular-ordering constraint on the phase space available
for the parton shower evolution. Thus, angular-ordered parton showers can
{\em consistently} compute the first two dominant towers ($\as^n L^{2n}$ and 
$\as^n L^{2n-1}$) of logarithmic contributions in Eq.~(\ref{logxs}).
However, parton showers contain also some subleading logarithmic contributions.
For instance, they correctly compute the single-logarithmic terms $\as^n L^n$ 
of purely collinear origin that lead to the LO evolution of the parton 
densities. 
Moreover, as discussed in Ref.~[\ref{CMW}] by a comparison with 
resummed calculations,
in the case of hard-scattering processes whose LO subprocess involves 
two coloured partons (e.g. DIS or DY production), angular-ordered parton showers
have a higher logarithmic accuracy: they can
consistently evaluate the LL and NLL terms in Eq.~(\ref{resff}).
The extension of parton-shower algorithms to higher logarithmic 
accuracy is not necessarily feasible and is, in any case, challenging. 

Of course, because of
quantum interferences and quantum fluctuations, the probabilistic parton-shower
approach cannot be used to systematically perform exact calculations at NLO,
NNLO and so forth. Nonetheless, important progress has been made to include
matrix element corrections in parton shower algorithms 
[\ref{mecorsey}--\ref{mecorcol}]. The purpose is to consider the multiparton
configurations generated by parton showering from the LO matrix element and to
correct them in the hard (non-soft and non-collinear) region
by using the exact expressions of the higher-order matrix elements.
Hard matrix element corrections to parton showers have been implemented
for top quark decay [\ref{seytop}] and for production of $W,Z$ and 
DY lepton pairs [\ref{sjow}, \ref{mrenna}, \ref{corseyw}]. The same techniques
could be applied to other processes, as, for instance, production of Higgs 
boson [\ref{resvsps}] and vector-boson pairs [\ref{proclhc}].

Note also that, at present, angular-ordered parton showers 
cannot be considered as true `next-to-leading' tools, even where their
logarithmic accuracy is concerned. 
The consistent computation of the first two towers of
logarithmic contributions in Eq.~(\ref{logxs}) is not sufficient for this
purpose. For instance, to precisely introduce an NLO definition of
$\as$, we should control all the terms obtained by the replacement $\as \to
\as + c \;\as^2 + {\cal O}(\as^3)$. When it is introduced in the towers
of double-logarithmic terms $\as^n L^{2n}$ of Eq.~(\ref{logxs}), 
this replacement leads to contributions of the type $\as^{n+1} L^{2n}
\sim \as^{n} L^{2n-2}$. Since these contributions are not
fully computable at present, the parameter $\as$ used in the parton showers
corresponds to a simple LO parametrization of QCD running 
coupling.

\section{Soft-gluon resummation}
\label{secsoftg}

Double-logarithmic contributions due to soft gluons arise in all the 
kinematic configurations where radiation of real and virtual partons
is highly unbalanced (see Ref.~[\ref{softrev}] and referenes therein).
For instance, this happens in the case of transverse-momentum distributions
at low transverse momentum, in the case of hard-scattering production near
threshold or when the structure of the final state
is investigated with high resolution (internal jet structure, shape variables).

Soft-gluon resummation for jet shapes has been extensively
studied and applied to hadronic final states produced by $e^+e^-$ annihilation
[\ref{eps99}, \ref{lp97}, \ref{jetres}]. Applications to hadron--hadron
collisions have just begun to appear [\ref{seymour}]
and have a large, 
yet uncovered, potential (from $\as$ determinations to studies of
non-perturbative dynamics).

Transverse-momentum logarithms, $L = \ln Q^2/{\bom Q}_{\perp}^2$,
occur in the distribution of transverse momentum ${\bom Q}_{\perp}$
of systems with high mass $Q$ $(Q \gg Q_{\perp})$ that are produced with
a vanishing ${\bom Q}_{\perp}$ in the LO subprocess. Examples of such systems
are DY lepton pairs, lepton pairs produced by $W$ and $Z$ decay, heavy 
quark--antiquark pairs, photon pairs and Higgs bosons. In these processes the LO 
transverse-momentum distribution is sharply peaked around ${\bom Q}_{\perp}=0$
($d {\hat \sigma}/d^2{\bom Q}_{\perp} \propto \delta^{(2)}({\bom Q}_{\perp}$)).
If the heavy system is produced with ${\bom Q}_{\perp}^2 \ll Q$, the emission
of real radiation at higher orders is strongly suppressed and cannot balance 
the virtual contributions. The ensuing logarithms, 
$L = \ln Q^2/{\bom Q}_{\perp}^2$, diverge order by order when 
${\bom Q}_{\perp} \to 0$, but after all-order resummation they leads to a 
finite smearing of the LO distribution.

Threshold logarithms, $L = \ln (1-x)$, occur when the tagged final state
produced by the hard scattering is forced to carry a very large fraction
$x$ ($x \to 1$) of the available centre-of-mass energy $\sqrt S$. Also in
this case, the radiative tail of real emission is stronly suppressed
at higher perturbative orders. Oustanding examples of hard processes near
threshold are DIS at large $x$ (here $x$ is the Bjorken variable), production
of DY lepton pairs with large invariant mass $Q$ ($x = Q/{\sqrt S}$), 
production of heavy quark--antiquark pairs ($x = 2m_Q/{\sqrt S}$), production
of single jets and single photons at large transverse energy $E_T$ 
($x = 2E_T/{\sqrt S}$).

To emphasize the difference between transverse-momentum logarithms and
threshold logarithms generated by soft gluons, it can be instructive to 
consider prompt-photon production. In the case of production of a {\em
photon pair}\footnote{The same discussion applies to the production of a DY
lepton pair.} with invariant mass squared 
$Q^2= (p^{(\gamma)}_{1}+p^{(\gamma)}_{2})^2$
and {\em total} transverse momentum 
${\bom Q}_{\perp}= {\bom p}^{(\gamma)}_{1 \perp}+{\bom p}^{(\gamma)}_{2 \perp}$,
transverse-momentum logarithms and threshold logarithms
appear when ${\bom Q}_{\perp}^2 \ll Q^2$ and ${\bom Q}_{\perp}^2 \sim
(S/4 - Q^2)$, respectively. However, in the case of production of a {\em
single photon} with transverse energy (or, equivalently, transverse momentum)
$E_T$, soft gluons can produce logarithms
only in the threshold  region $x_T = 2E_T/{\sqrt S} \to 1$. If the prompt
photon has a transverse energy that is not close\footnote{Eventually,
when $x_T \ll 1$, higher-order corrections are single-logarithmically
enhanced. This small-$x$ logarithms, $(\as \ln x_T)^n$, have to be taken into
account by BFKL resummation.}
to its threshold value, 
the emission of accompanying radiation is not kinematically suppressed
and there are no soft logarithms analogous to those in the
transverse-momentum distribution of a photon pair. In particular, 
there are no double-logarithmic contributions of the type 
$(\as \ln^2 E_T^2/S)^n$, and perturbative soft gluons are not distinguishable
from perturbative hard gluons.

Studies of soft-gluon resummation for transverse-momentum distributions
at low transverse momentum and for hard-scattering production near
threshold started two decades ago [\ref{BCM}, \ref{DDT}]. The physical bases
for a systematic all-order summation of the soft-gluon contributions are
dynamics and kinematics factorizations [\ref{softrev}]. The first factorization
follows from gauge invariance and unitarity: in the soft limit, multigluon
amplitudes fulfil factorization formulae given in terms of universal
(process-independent) soft contributions. The second factorization regards 
kinematics and strongly depends on the actual cross section to be evaluated.
{\em If}, in the appropriate soft limit, the multiparton phase space for
this cross section can be written in a factorized way, resummation
is analytically feasible in form of {\em generalized exponentiation}
of the universal soft contributions that appear in the factorization formulae
of the QCD amplitudes.

Note that the phase space depends in a 
non-trivial way on multigluon configurations and, in general, is not
factorizable in single-particle contributions\footnote{In the case of jet
cross sections, for instance, phase-space factorization depends on the
detailed definition of jets and it can easily be violated [\ref{BS}]. Some
jet algorithms, such as the $k_{\perp}$-algorithm 
[\ref{duralg}, \ref{ktalg}], have better
factorization properties.}.
Moreover, even when phase-space factorization is
achievable, it does not always occur in the space of the kinematic variables
where the cross section is defined. Usually, it is necessary to
introduce a conjugate space to overcome phase-space constraints. This is the
case for transverse-momentum distributions and hard-scattering production near
threshold. The relevant kinematical constraint for 
${\bom Q}_{\perp}$-distributions is (two-dimensional) 
transverse-momentum conservation and it
can be factorized by performing a Fourier transformation. Soft-gluon 
resummation for ${\bom Q}_{\perp}$-distributions is thus carried out
in ${\bom b}$-space [\ref{pp}, \ref{CSS}], where the impact parameter 
${\bom b}$ is the variable conjugate to ${\bom Q}_{\perp}$ via the Fourier
transformation. Analogously, the relevant kinematical constraint for
hard-scattering production near threshold is (one-dimensional) energy
conservation and it can be factorized by working in $N$-moment space 
[\ref{S}, \ref{CT}], 
$N$ being the variable conjugate to the threshold variable $x$ 
(energy fraction) via a Mellin or Laplace transformation.

Using a short-hand notation, 
the general structure of the partonic cross section ${\hat \sigma}$
after summation of soft-gluon contributions is
\beq
\label{resmatch}
{\hat \sigma} = {\hat \sigma}_{\rm res.} + {\hat \sigma}_{\rm rem.} \;.
\eeq
The term ${\hat \sigma}_{\rm res.}$ embodies the all-order resummation,
while the remainder ${\hat \sigma}_{\rm rem.}$ contains no large logarithmic
contributions. The latter has the form
\beq
{\hat \sigma}_{\rm rem.} = {\hat \sigma}^{({\rm f.o.})} -
\left[ \,{\hat \sigma}_{\rm res.} \,\right]^{({\rm f.o.})} \;\;,
\eeq
and it is obtained from ${\hat \sigma}^{({\rm f.o.})}$,
the truncation of the perturbative expansion for ${\hat \sigma}$ at a given
fixed order (LO, NLO, ...), by subtracting the corresponding truncation
$\left[ {\hat \sigma}_{\rm res.}\right]^{({\rm f.o.})}$ of the resummed part.
Thus, the expression on the right-hand side of Eq.~(\ref{resmatch}) includes
soft-gluon logarithms to all orders and it is {\it matched} to the exact (with
no logarithmic approximation) fixed-order calculation. It represents an
improved perturbative calculation that is everywhere as good as the fixed-order
result, and much better in the kinematics regions where the soft-gluon
logarithms become large ($\as L \sim 1$). Eventually, when $\as L \gtap 1$, the
resummed perturbative contributions are of the same size as the 
non-perturbative contributions and the effect of the latter has to be 
implemented in the resummed calculation.

The resummed cross section has the following typical form:
\beq
{\hat \sigma}_{\rm res.} = 
\as^k \int_{\rm inv.} {\hat \sigma}^{(LO)} \cdot C \cdot S \;,
\eeq
where the integral $\int_{\rm inv.}$ denotes the inverse tranformation from
the conjugate space where resummation is actually carried out.
Methods to perform the inverse transformation are discussed in 
Refs.~[\ref{qtback}] and [\ref{thback}] for $Q_{\perp}$-resummation and
threshold resummation, respectively. The $C$ term has the perturbative
expansion
\beq
C = 
1 + C_1 \,\as + C_2 \,\as^2 + \dots \,
\eeq
and contains all the constant contributions in the limit $L \to \infty$
(the coefficients $C_1, C_2, \dots$ do not depend on the conjugate variable).
The singular dependence on $L$ (more precisely, on the logarithm ${\tilde L}$
of the conjugate variable) is entirely {\it exponentiated} in the
factor $S$: 
\beq
\label{resff}
S = \exp \left\{ L \;g_1(\as L) + g_2(\as L) +
\as \;g_3(\as L) + \dots \right\} \;\;.
\eeq
In the exponent, the function $L \,g_1$
resums all the leading logarithmic (LL) contributions
$\as^n L^{n+1}$, while $g_2$
contains the next-to-leading logarithmic (NLL) terms $\as^n L^n$ and so 
forth\footnote{To compare this notation with that of Ref.~[\ref{qthere}], we 
can notice that our functions $g_i$ are obtained by the straightforward
integration over ${\overline \mu}$ of the functions $A(\as({\overline \mu}))$ 
and $B(\as({\overline \mu}))$ of Ref.~[\ref{qthere}]. In particular, our
terms $g_1, g_2, g_3$ are not to be confused with the non-perturbative
parameters of the same name used in Ref.~[\ref{qthere}].}
(all the functions $g_i$ are normalized as $g_i(\lambda=0) = 0$).
Note that the LL terms are formally suppressed by a power of $\as$ with respect
to the NLL terms, and so forth for the successive classes of logarithmic terms. 
Thus, this logarithmic expansion is as systematic
as the fixed-order expansion in Eq.~(\ref{pertex}).
In particular, using a matched NLL+NLO calculation, we can 
consistently $i)$ introduce a precise definition (say ${\overline {\rm MS}}$)
of $\as(\mu)$ and $ii)$ investigate the theoretical accuracy of the calculation
by studying its renormalization-scale dependence. 

The structure of the exponentiated resummed calculations discussed so far
has to be contrasted with that obtained by organizing the logarithmic expansion
on the right-hand side of Eq.~(\ref{logxs}) in terms of towers as 
\beq
\label{restowerex}
{\hat \sigma} \sim \as^k {\hat \sigma}^{(LO)} 
\left\{ t_1(\as L^2) + \as L \; t_2(\as L^2) + \as^2 L^2 \; t_3(\as L^2)
+ \dots  \right\} \;\;,
\eeq
where the double-logarithmic function $t_1(\as L^2)$ and the successive
functions are normalized as $t_i(0)= {\rm const.}$ While the ratio of two
successive terms in the exponent of Eq.~(\ref{resff}) is formally of the order
of $\as$, the ratio of two successive towers in Eq.~(\ref{restowerex}) is 
formally of the order of $\as L$. In other words, the tower expansion sums the
double-logarithmic terms $(\as L^2)^n$, then the terms $\as^n L^{2n-1} \sim 
\as L (\as L^2)^{n-1}$, and so forth; it thus 
assumes that the resummation procedure is carried out with respect to the 
large parameter $\as L^2$ ($\as L^2 \ltap 1$). On the contrary, in 
Eq.~(\ref{resff}) the large parameter is $\as L \ltap 1$. The tower expansion
allows us to formally extend the applicability of perturbative QCD to the
region $L \ltap 1/{\sqrt \as}$, and exponentiation extends it to the wider
region $L \ltap 1/\as$. This fact can also be argued by comparing the
amount of information on the logarithmic terms that is included in the
truncation of Eqs.~(\ref{resff}) and (\ref{restowerex}) at some logarithmic
accuracy. The reader can easily check that, after matching to the NLO (LO)
calculation as in Eq.~(\ref{resmatch}), the NLL (LL) result of Eq.~(\ref{resff})
contains all the logarithms of the first {\em four} ({\em two}) towers in
Eq.~(\ref{restowerex}) (and many more logarithmic terms).

In the case of $Q_{\perp}$-distributions, full NLL resummation has been 
performed for lepton pairs, $W$ and $Z$ bosons produced by the DY mechanism
[\ref{CSS}, \ref{dyqt}] and for Higgs bosons produced by gluon fusion
[\ref{Hqt}]. Corresponding resummed calculations are discussed in 
Refs.~[\ref{resvsps}, \ref{qthere}] and references therein.

Threshold logarithms in hadron collisions have been resummed to NLL
accuracy for DIS and DY production [\ref{S}, \ref{CT}, \ref{CMW}, \ref{CLS}]
and for Higgs boson production [\ref{Hth}].
Recent theoretical progress [\ref{KS}, \ref{KOS}, \ref{txs}]
regards the extension of NLL resummation
to processes produced by LO hard-scattering of more than two coloured partons,
such as heavy-quark hadroproduction [\ref{KS}, \ref{txs}] and leptoproduction
[\ref{LM}], as well as prompt-photon [\ref{LOS}--\ref{kow}],
quarkonium [\ref{cacciari}] and vector-boson [\ref{kdd}] production.

An important feature of threshold resummation
is that the resummed soft-gluon contributions regard the partonic cross
section rather than the hadronic cross section. This fact has two main
consequences: $i)$ soft-gluon contributions can be sizeable long before the 
threshold region in the hadronic cross section is actually approached, and 
$ii)$ the resummation effects typically enhance the fixed-order perturbative
calculations. 

The first consequence follows from the fact that the evolution of the parton
densities sizeably reduces the energy that is available in the 
partonic hard-scattering subprocess. Thus, the partonic cross section 
$\hat \sigma$ in the factorization formula
(\ref{factfor}) is typically evaluated much closer to threshold than the 
hadronic cross section. In other words, the
parton densities are strongly suppressed at large $x$ (typically, 
when $x\to 1$, $f(x,\mu^2) \sim (1-x)^\eta$ with $\eta \sim 3$ and 
$\eta \sim 6$ for valence quarks and sea-quarks or gluons, respectively); 
after integration over them, the dominant value of the square of the
partonic centre-of-mass energy $\langle {\hat s} \rangle = 
\langle x_1 x_2 \rangle S$ is therefore
substantially smaller than the corresponding hadronic value $S$.

The second consequence, which depends on the actual definition of the parton 
densities,
follows from the fact that the resummed contributions
are those soft-gluon effects that are left at the partonic level after 
factorization of the parton densities. After having absorbed part of the
full soft-gluon contributions in the customary definitions (for instance, those
in the ${\overline {\rm MS}}$ or DIS factorization schemes) of the parton 
densities, it turns out that the residual effect in the partonic cross section 
is positive and tends to enhance the perturbative predictions.

A quantitative illustration of these consequences is given below by discussing
top-quark and prompt-photon production. The discussion also shows another
relevant feature of NLO+NLL calculations, namely, their increased stability
with respect to scale variations.

\begin{figure}
  \centerline{
    \setlength{\unitlength}{1cm}
    \begin{picture}(0,6)
       \put(0,0){\includegraphics{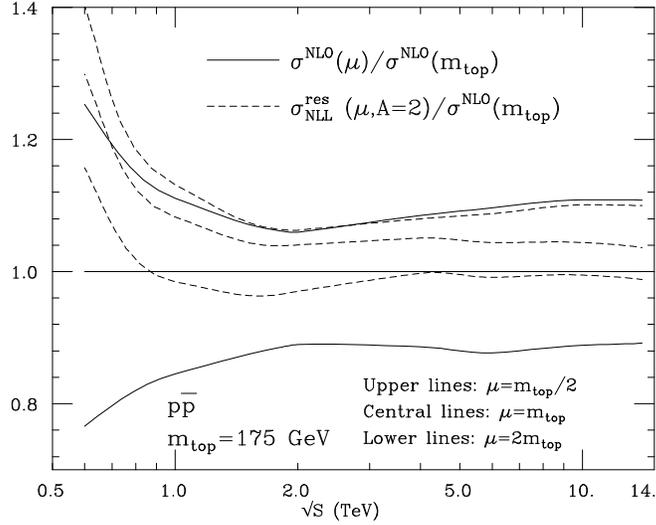}}
    \end{picture}}
\vspace*{1.3cm}    
\caption{The $t{\bar t}$ production cross section in $p{\bar p}$ collisions as a
function of ${\sqrt S}$. The solid lines represent the NLO results for different
choices ($\mu=m_t/2$ and $\mu=2m_t$) of the renormalization/factorization 
scale $\mu=\mu_R=\mu_F$, normalized to the result with $\mu=m_t$. The dashed
lines represent the NLO+NLL results for different choices of $\mu$
($\mu=m_t/2, m_t$ and $2m_t$), normalized to the NLO result with 
$\mu=m_t$.
\label{figtop}}
\end{figure}

The effects of soft-gluon resummation on the top-quark production cross sections
at hadron colliders have been studied in Refs.~[\ref{txs}, 
\ref{tqll1}--\ref{tqll4}]. In the case of $p{\bar p}$ collisions, the 
comparison between QCD predictions at NLO and those after NLL resummation is shown in
Fig.~\ref{figtop} [\ref{txs}]. 
At the Tevatron the resummation effects are not very large 
and the NLO cross section is increased by about $4\%$. This had to be expected
because the top quark is not produced very close
to threshold ($x = 2m_t/{\sqrt S} \sim 0.2$, at the Tevatron). Note, however,
that the dependence on the factorization/renormalization scale of the
theoretical cross section is reduced by a factor of  almost 2 by including NLL
resummation. More precisely, the scale dependence $(\sim \pm 5\%)$
of the NLO+NLL calculation becomes comparable to that obtained by using 
different sets of parton densities [\ref{pdffit}]. Combining linearly scale
and parton density uncertainties, the NLO+NLL cross section is
$\sigma_{t{\bar t}}=5.0 \pm 0.6$, with $m_t=175~{\rm GeV}$ and 
${\sqrt S}=1.8~{\rm TeV}$ [\ref{txs}]. 

At the LHC $(x = 2m_t/{\sqrt S} \sim 0.03)$ the top quark is produced less close
to the hadronic threshold than at the Tevatron. However this is compensated
by the fact that the gluon channel\footnote{Since
$f_g$ is steeper than $f_q$ at large $x$, partonic cross sections in gluon
subprocesses are typically closer to threshold than in quark subprocesses.
Moreover, the intensity of soft-gluon radiation from gluons is larger than 
that from
quarks by a factor of $\sim C_A/C_F \sim 2$.} is more important at the LHC.
As a result, the effect of including
soft-gluon resummation to NLL accuracy is very similar: 
the NLO cross section is enhanced by $\sim 5\%$ and its scale dependence
is reduced from $\sim \pm 10\%$ to $\sim \pm 5\%$.
Note, however, that the uncertainty $(\sim \pm 10\%)$ coming from the parton
(gluon) densities is larger than at the Tevatron [\ref{proclhc}]. 

\begin{figure}
  \centerline{
    \setlength{\unitlength}{1cm}
    \begin{picture}(0,6)
       \put(0,0){\includegraphics{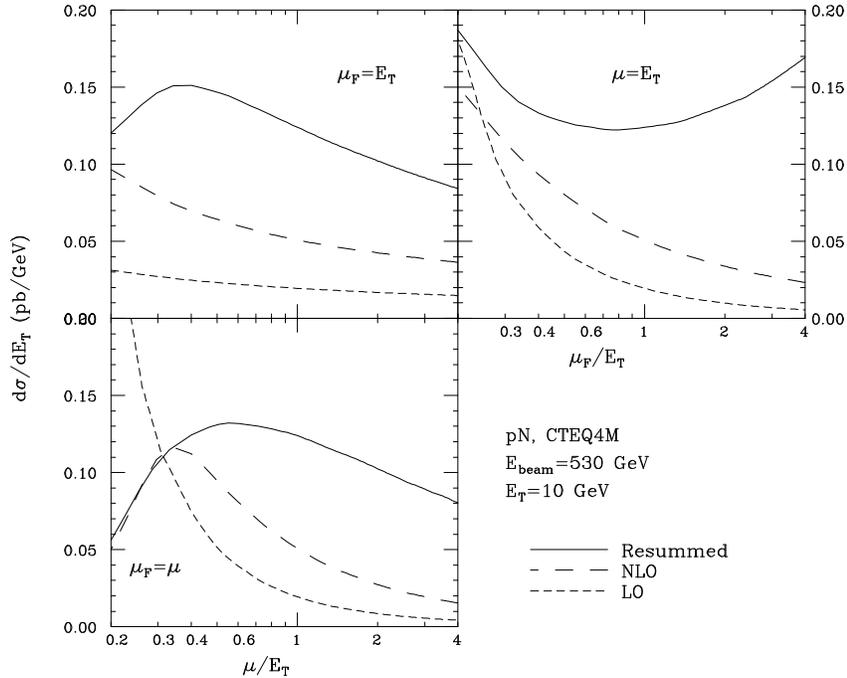}}
    \end{picture}}
\vspace*{3.0cm}    
\caption{The dependence on the factorization $(\mu_F)$ and renormalization
$(\mu_R=\mu)$ scale of the prompt-photon cross section 
$d\sigma/dE_T$ in $pN$ collisions at $E_T=10$~GeV and ${\sqrt S}=31.6$~GeV. 
The short-dashed, long-dashed and solid lines are respectively the results at
LO, NLO and after NLO+NLL resummation. 
\label{figgamma10}}
\end{figure}

\begin{figure}
  \centerline{
    \setlength{\unitlength}{1cm}
    \begin{picture}(0,6)
       \put(0,0){\includegraphics{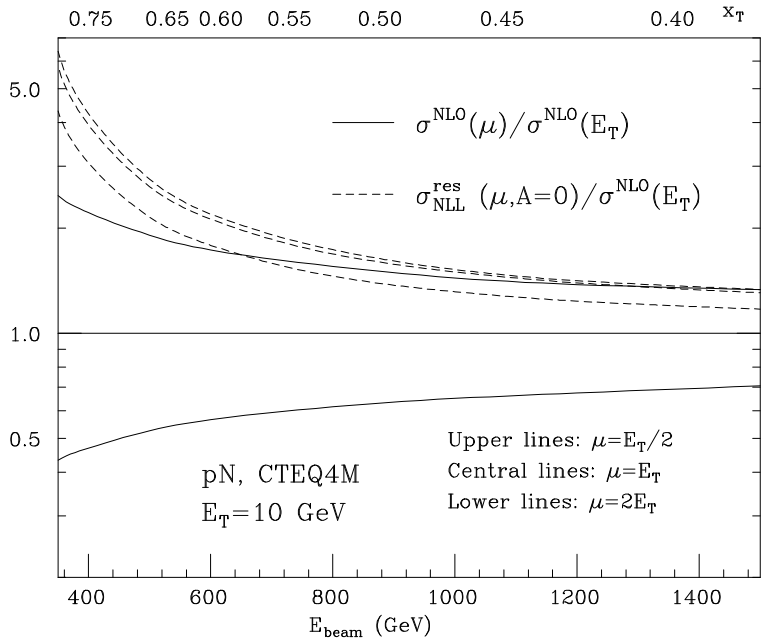}}
    \end{picture}}
\vspace*{2.5cm}    
\caption{The prompt-photon cross section $d\sigma/dE_T$ in $pN$ collisions 
at $E_T=10$~GeV as a function of the energy $E_{{\rm beam}}$
of the proton beam. The solid lines represent the NLO results for 
different choices ($\mu=E_T/2$ and $\mu=2E_T$) of the 
renormalization/factorization 
scale $\mu=\mu_R=\mu_F$, normalized to the result with $\mu=E_T$. The dashed
lines represent the NLO+NLL results for different choices of $\mu$
($\mu=E_T/2, E_T$ and $2E_T$), normalized to the NLO result with $\mu=E_T$.
\label{figgammabeam}}
\end{figure}

Similar qualitative results are obtained [\ref{CMNOV}] when NLL resummation
is applied to prompt-photon production at fixed-target experiments. The scale
dependence of the theoretical calculation is highly reduced and the resummed
NLL contributions lead to large corrections at high $x_T=2E_T/{\sqrt S}$
(and smaller corrections at lower $x_T$). Of course, the impact of
soft-gluon resummation is quantitatively more sizeable in 
prompt-photon production than in top-quark production, because $x_T$
can be as large as 0.6, the hard scale $E_T$ is much smaller than $m_t$
(thus, $\as(E_T) > \as(m_t)$) and the gluon channel is always important. The
scale dependence of the theoretical cross section for the E706 kinematics
is shown in Fig.~\ref{figgamma10}. Fixing $\mu_R=\mu_F=\mu$ and varying $\mu$
in the range $E_T/2 < \mu < 2E_T$ with $E_T=10~{\rm GeV}$, the cross section
varies by a factor of $\sim 6$ at LO, by a factor of $\sim 4$ at NLO and
by a factor of $\sim 1.3$ after NLL resummation. The highly reduced scale
dependence of the NLO+NLL cross section is also visible in 
Fig.~\ref{figgammabeam}, which, in particular, shows that when 
$E_T=10~{\rm GeV}$ and $E_{\rm beam}=530~{\rm GeV}$ the central value (i.e. with
$\mu=E_T$) of the NLO cross section increases by a factor of $\sim 2.5$ after
NLL resummation. As expected, the size of these effects is reduced by increasing
${\sqrt S}$ at fixed $E_T$ (see Fig.~\ref{figgammabeam}) or by decreasing 
$E_T$ at fixed ${\sqrt S}$ (see Fig.~\ref{figgamma5}).

\begin{figure}
  \centerline{
    \setlength{\unitlength}{1cm}
    \begin{picture}(0,6)
       \put(0,0){\includegraphics{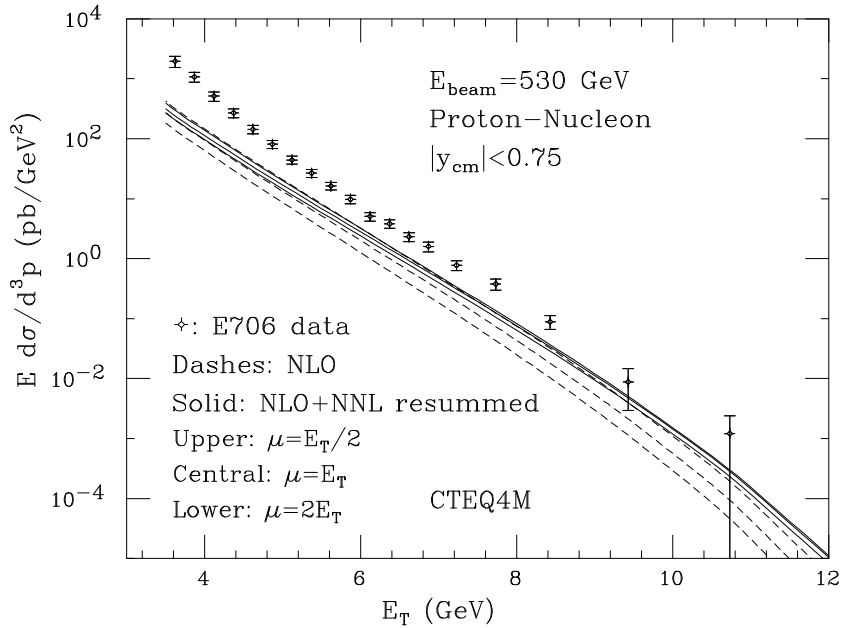}}
    \end{picture}}
\vspace*{2.7cm}    
\caption{E706 prompt-photon data compared with theoretical calculations, 
which use the parton
densities of the set CTEQ4 and GRV photon fragmentation functions.
The solid and dashed lines correspond to the NLO+NLL and pure NLO calculations,
respectively.
\label{fige706vsres}}
\end{figure}

The comparison with the E706 data shown in Fig.~\ref{fige706vsres} suggests
that the NLO+NLL calculation can help to
better understand prompt-photon production at large $x_T$. Note, however, that
this comparison has to be regarded as preliminary in several respects 
[\ref{CMNOV}]. In particular, the parton densities used in 
Fig.~\ref{fige706vsres} are those extracted from NLO fits. Owing to the
soft-gluon enhancement at large $x_T$, refitting the parton densities may lead
to a smaller $f_g$ at large $x$ and, consequently (because of the momentum sum
rule), a larger $f_g$  at intermediate $x$. As a result, this procedure could
somehow increase the theoretical cross section also at smaller values of $x_T$.

Soft-gluon resummation at NLL accuracy is now available for all the processes
(namely, DIS, 
DY 
and prompt-photon 
production)
that are typically used to perform global fits to parton densities.
A detailed extraction/evolution of parton densities by consistently using
NLL resummed calculations is thus nowadays feasible.

\section{Other topics}

The activity of the QCD Working Group at this Workshop has also been devoted 
to other topics, such as
automatic computation of matrix elements and LO cross sections for
multiparticle processes at high-energy colliders, 
definition and properties of jets algorithms, 
definition of isolated photons and related NLO calculations. 
Corresponding contributions are included in these Proceedings. 

Other studies performed during this Workshop have a large overlap
with the activity of the related Workshops at FERMILAB and CERN 
and can be found in those Proceedings [\ref{proctev}, \ref{proclhc}].

\noindent {\bf Acknowledgements}. I would like to thank the members of the
Local Organizing Committee for the excellent Workshop. 

\section*{References}

\def\ac#1#2#3{Acta Phys.\ Polon.\ #1 (19#3) #2}
\def\ap#1#2#3{Ann.\ Phys.\ (NY) #1 (19#3) #2}
\def\ar#1#2#3{Annu.\ Rev.\ Nucl.\ Part.\ Sci.\ #1 (19#3) #2}
\def\cpc#1#2#3{Computer Phys.\ Comm.\ #1 (19#3) #2}
\def\ib#1#2#3{ibid.\ #1 (19#3) #2}
\def\np#1#2#3{Nucl.\ Phys.\ B#1 (19#3) #2}
\def\pl#1#2#3{Phys.\ Lett.\ #1B (19#3) #2}
\def\pr#1#2#3{Phys.\ Rev.\ D #1 (19#3) #2}
\def\prep#1#2#3{Phys.\ Rep.\ #1 (19#3) #2}
\def\prl#1#2#3{Phys.\ Rev.\ Lett.\ #1 (19#3) #2}
\def\rmp#1#2#3{Rev.\ Mod.\ Phys.\ #1 (19#3) #2}
\def\sj#1#2#3{Sov.\ J.\ Nucl.\ Phys.\ #1 (19#3) #2}
\def\zp#1#2#3{Z.\ Phys.\ C#1 (19#3) #2}
\def\epj#1#2#3{Eur.\ Phys.\ J. C#1 (19#3) #2}
\def\jhep#1#2#3{JHEP #1 (19#3) #2}

\begin{enumerate}

\item \label{eps99}
M.L.\ Mangano, plenary talk at the {\it 1999 International Europhysics 
Conference on High Energy Physics} Tampere, Finland, 15--21 July 1999, preprint
CERN-TH/99-337 (hep-ph/9911256), to appear in the Proceedings, 
and references therein.

\item \label{lp99webber}
B.R.\ Webber, plenary talk at the {\it 19th International Symposium on 
Lepton-Photon Interactions at High-Energies, LP 99}, Stanford,
California, 9--14 Aug 1999, preprint CERN-TH/99-387
(hep-ph/9912292), to appear in the Proceedings, 
and references therein. 

\item \label{ichep98}
J.\ Huston, in Proc. of the {\it 29th International Conference on High-Energy
Physics, ICHEP 98}, eds. A.~Astbury, D.~Axen and J. Robinson
(World Scientific, Singapore, 1999), vol.~1, p.~283, and references therein.

\item \label{lp97}
S.\ Catani, hep-ph/9712442, in Proc. of the {\it XVIII International Symposium
on Lepton-Photon Interactions, LP97}, eds. A.\ De Roeck and A.\ Wagner
(World Scientific, Singapore, 1998), p.~147,
and references therein.

\item \label{proceeding}
G.\ Jarlskog and D.\ Rein eds., Proc. of ECFA LHC Workshop, report CERN 90-10
(December 1990);
ATLAS Coll., ATLAS TDR 15, report CERN/LHCC/99-15 (May 1999).

\item \label{proctev}
Proceedings of the Workshop on {\it Physics at the Tevatron in Run II}, Fermilab,
2000 (to appear).
See: http://www-theory.fnal.gov/people/ellis/QCDWB/QCDWB.html

\item \label{proclhc}
Proceedings of the Workshop on {\it Standard Model Physics (and more) at the LHC},
CERN 1999 (to appear). 
See: http://home.cern.ch/\~{}mlm/lhc99/lhcworkshop.html

\item \label{factform}
See J.C.\ Collins, D.E.\ Soper and G.\ Sterman, in {\it Perturbative Quantum
Chromodynamics}, ed. A.H.\ Mueller (World Scientific, Singapore, 1989), p.~1,
and references therein.

\item \label{DGLAP}
V.N.\ Gribov and L.N.\ Lipatov, Sov. J. Nucl. Phys. 15 (1972) 438, 
675; G.\ Altarelli and G.\ Parisi,
\np{126}{298}{77}; Yu.L.\ Dokshitzer, Sov. Phys. JETP  46 (1977) 641.

\item \label{NLOAP}
E.~G.~Floratos, D.~A.~Ross and C.~T.~Sachrajda,
Nucl.\ Phys.\  B129 (1977) 66, E ibid. B139 (1978) 545,
Nucl.\ Phys.\  B152 (1979) 493;
A.~Gonzalez-Arroyo, C.~Lopez and F.~J.~Yndurain,
Nucl.\ Phys.\  B153 (1979) 161;
A.~Gonzalez-Arroyo and C.~Lopez, Nucl.\ Phys.\  B166 (1980) 429;
G.~Curci, W.~Furmanski and R.~Petronzio, Nucl.\ Phys.\  B175 (1980) 27;
W.~Furmanski and R.~Petronzio, Phys.\ Lett.\  B97 (1980) 437;
E.~G.~Floratos, C.~Kounnas and R.~Lacaze, Nucl.\ Phys.\  B192 (1981) 417.
 
\item \label{pdffit}
A.D.\ Martin, R.G.\ Roberts, W.J.\ Stirling and R.S. Thorne, \epj{4}{463}{98},
\pl{443}{301}{98},
preprint DTP-99-64 (hep-ph/9907231); 
M.\ Gl\"uck, E.\ Reya and A. Vogt, \epj{5}{461}{98}; 
H.L. Lai et al.,
Eur.\ Phys.\ J. C12 (2000) 375.

\item \label{lhcpdf}
LHC Guide to Parton Distribution Functions and Cross Sections, ATLAS note
ATL-PHYS-99-008, http://www.pa.msu.edu/\~{}huston/lhc/lhc$_{-}$pdfnote.ps .

\item \label{pdfrball}
R.D.~Ball, in these Proceedings, and references therein.

\item \label{bodek}
U.K.~Yang and A.~Bodek, \prl{82}{2467}{99},
preprint UR-1581 (hep-ex/9908058).

\item \label{Melnitchouk}
W.~Melnitchouk, I.R.~Afnan, F.~Bissey and A.W.~Thomas,
preprint ADP-99-47-T384 (hep-ex/9912001). 

\item \label{highxcteq}
S. Kuhlmann et al., hep-ph/9912283.

\item \label{gluoncteq}
J. Huston et al., \pr{58}{114034}{98}. 

\item \label{NNLODY}
R.\ Hamberg, W.L.\ van Neerven and T.\ Matsuura, \np{359}{343}{91};
W.L.~van Neerven and E.B.\ Zijlstra, \np{382}{11}{92}.

\item \label{txs}
R.\ Bonciani, S.\ Catani, M.L.\ Mangano and P.\ Nason, \np{529}{424}{98}.

\item \label{BFKL}
L.N.\ Lipatov, Sov. J. Nucl. Phys. 23 (1976) 338; E.A.\ Kuraev,
L.N.\ Lipatov and V.S.\ Fadin, Sov. Phys. JETP  45 (1977) 199; Ya.\
Balitskii and L.N.\ Lipatov, Sov. J. Nucl. Phys. 28 (1978) 822.

\item \label{Jar}
T.\ Jaroszewicz, \pl{116}{291}{82}.

\item \label{CH}
S.\ Catani and F.\ Hautmann, \pl{315}{157}{93}, \np{427}{475}{94}.

\item \label{fadin}
V.S.\ Fadin and L.N.\ Lipatov, \sj{50}{712}{89}, \pl{429}{127}{98}.
 
\item \label{CC}
M. Ciafaloni and G. Camici, 
\np{496}{305}{97}, \pl{430}{349}{98}. 

\item \label{sdis}
S.\ Catani, \zp{70}{263}{96}, \zp{75}{665}{97}.

\item \label{hqnlo}
P.\ Nason, S.\ Dawson and R.K.\ Ellis, \np{303}{607}{88};
W.\ Beenakker, H.~Kuijf, W.L.\ van Neerven and J.\ Smith, \pr{40}{54}{89};
M.L.\ Mangano, P.\ Nason and G.\ Ridolfi, \np{373}{295}{92}.

\item \label{baarmand}
M.~Baarmand, talk presented at the
Workshop on {\it Standard Model Physics (and more) at the LHC},
CERN, January 1999. \\
See: http://home.cern.ch/n/nason/www/lhc99/14-01-99/program-1-14-99

\item \label{bquarkrev}
S.~Frixione, M.L.~Mangano, P.~Nason and G.~Ridolfi, in {\it Heavy Flavours II},
eds. A.J.~Buras and M. Lindner (World Scientific, Singapore, 1998), p.~609, and
references therein.

\item \label{ptbres}
M.\ Cacciari, M.\ Greco and P. Nason, \jhep{05}{007}{98};
F.I.\ Olness, R.J.\ Scalise and Wu-Ki Tung, \pr{59}{014506}{99}.

\item \label{ktfac}
S. Catani, M. Ciafaloni and F. Hautmann, \pl{242}{97}{90}, \np{366}{135}{91};
J.C. Collins and R.K. Ellis, \np{360}{3}{91};
E.M.\ Levin, M.G.\ Ryskin, Yu.M.\ Shabel'skii and A.G.\ Shuvaev, Sov. J.
Nucl. Phys. 53 (1991) 657.

\item \label{promptgnlo}
P.\ Aurenche, R.\ Baier, M.\ Fontannaz and D.\ Schiff, \np{297}{661}{88};
H.~Baer, J. Ohnemus and J.F. Owens, \pr{42}{61}{90};
P.\ Aurenche, R.~Baier and M.\ Fontannaz, \pr{42}{1440}{90};
L.E.\ Gordon and W.~Vogelsang, \pr{48}{3136}{93}. 

\item \label{annecygamma}
P.\ Aurenche, M.\ Fontannaz, J.Ph.\ Guillet, B.\ Kniehl,
E.\ Pilon and M. Werlen, \epj{9}{107}{99}.

\item \label{E706}
E706 Coll., L. Apanasevich et al., \prl{81}{2642}{98}.

\item \label{ktcteq}
L. Apanasevich et al., \pr{59}{074007}{99}.

\item \label{ktmartin}
M.A.~Kimber, A.D.~Martin and M.G.~Ryskin, Eur. Phys. J. C12 (2000) 655. 

\item \label{ktsterman}
E.~Laenen, G.~Sterman and W.~Vogelsang, preprint YITP-99-69
(hep-ph/0002078).

\item \label{annecypi}
P.~Aurenche, M.~Fontannaz, J.P.~Guillet, B.A.~Kniehl and M.~Werlen, 
preprint LAPTH-751-99 (hep-ph/9910252).

\item \label{jh}
J.\ Huston et al., paper in preparation.

\item \label{delduca}
V.~Del Duca and G.~Heinrich, in these Proceedings, and references therein.

\item \label{book}
R.K.\ Ellis, W.J.\ Stirling and B.R.\ Webber, {\it QCD and Collider 
Physics} (Cambridge University Press, Cambridge, 1996) and references therein.

\item \label{Dokbook}
Yu.L.\ Dokshitzer, V.A.\ Khoze, A.H.\ Mueller and S.I.\ Troian, 
{\it Basics of Perturbative QCD} (Editions Fronti\`eres, Gif-sur-Yvette, 1991)
and references therein.

\item \label{BCM}
A.\ Bassetto, M.\ Ciafaloni and G.\ Marchesini, \prep{100}{201}{83}, and
references therein.

\item \label{CMW}
S.\ Catani, G.\ Marchesini and B.R. Webber, \np{349}{635}{91}.

\item \label{mecorsey}
M.H.~Seymour, Comput. Phys. Commun. 90 (1995) 95. 

\item \label{mecorsjo}
J.~Andre and T.~Sjostrand, Phys. Rev. D57 (1998) 5767.

\item \label{mecorfri}
C.~Friberg and T.~Sjostrand, in Proc. of the Workshop
{\it Monte Carlo Generators for HERA Physics}, eds. T.A.~Doyle, G.~Grindhammer,
G.~Ingelman and H.~Jung, (DESY, Hamburg, 1999), p.~181. 

\item \label{mecorcol}
J.C.~Collins, hep-ph/0001040 and in these Proceedings.

\item \label{seytop}
G.~Corcella and M.H.~Seymour, \pl{442}{417}{98}.

\item \label{sjow}
G.~Miu and T.~Sjostrand, \pl{449}{313}{99}.

\item \label{mrenna}
S. Mrenna,  preprint UCD-99-4 (hep-ph/9902471).

\item \label{corseyw}
G.~Corcella and M.H.~Seymour, preprint RAL-TR-1999-051 (hep-ph/9908388).

\item \label{resvsps}
C.~Balazs, J.~Huston and I.~Puljak, 
preprint FERMILAB-PUB-00-032-T \\
(hep-ph/0002032), and in these Proceedings. 
 
\item \label{softrev}
G.\ Sterman, in Proc. {\it 10th Topical Workshop on Proton-Antiproton Collider
Physics}, eds. R.\ Raja and J.\ Yoh (AIP Press, New York, 1996), p.~608;
S.\ Catani, Nucl. Phys. Proc. Suppl. 54A (1997) 107, and
in Proc of the {\it 32nd Rencontres de Moriond: QCD and High-Energy
Hadronic Interactions}, ed. J.\ Tran Than Van (Editions Fronti\`eres, Paris,
1997), p.~331. 

\item \label{jetres}
S.\ Catani, G.\ Turnock, B.R.\ Webber and L.\ Trenta\-due, \np{407}{3}{93}.

\item \label{seymour}
M.H.\ Seymour, \np{513}{269}{98}; 
J.R.\ Forshaw and M.H.\ Seymour, \jhep{09}{009}{99}.  

\item \label{DDT}
Yu.L.\ Dokshitzer, D.I.\ Diakonov and S.I.\ Troian, Phys. Rep. 58 (1980) 269,
and references therein.

\item \label{BS}     
N.\ Brown and W.J.\ Stirling, Phys. Lett. 252B (1990) 657.

\item \label{duralg}
S. Catani, Yu.L.\ Dokshitzer, M.\ Olsson, G.\ Turnock and B.R. Webber,
\pl{269}{432}{91}.

\item \label{ktalg}
S.\ Catani, Yu.L.\ Dokshitzer and B.R. Webber, \pl{285}{291}{92};
S.\ Catani, Yu.L.\ Dokshitzer, M.H.\ Seymour and B.R. Webber, \np{406}{187}{93};
S.D.~Ellis and D.E.\ Soper, \pr{48}{3160}{93}. 

\item \label{pp}
G.\ Parisi and R.\ Petronzio, \np{154}{427}{79}.  

\item \label{CSS}
J.C.\ Collins, D.E.\ Soper and G.\ Sterman, \np{250}{199}{85}.

\item \label{S}
G.\ Sterman, \np{281}{310}{87}.

\item \label{CT}
S.\ Catani and L.\ Trentadue, \np{327}{323}{89}, \np{353}{183}{91}.

\item \label{qtback}
J.C.\ Collins and D.E.\ Soper, \np{197}{446}{82}.

\item \label{thback}
S.\ Catani, M.L.\ Mangano, P.\ Nason and L.\ Trentadue,
\np{478}{273}{96}. 

\item \label{dyqt}
J.\ Kodaira and L. Trentadue, \pl{112}{66}{82},
\pl{123}{335}{83};
C.T.H.\ Davies, B.R.\ Webber and W.J. Stirling, \np{256}{413}{85}. 

\item \label{Hqt}
S.\ Catani, E.\ D'Emilio and L.\ Trentadue, \pl{211}{335}{88};
R.P.\ Kauffman, \pr{45}{1512}{92}. 

\item \label{qthere}
C.~Bal\'azs, J.C.~Collins and D.E.~Soper, in these Proceedings.

\item \label{CLS}
H.\ Contopanagos, E.\ Laenen and G. Sterman, \np{484}{303}{97}.

\item \label{Hth}
M. Kramer, E. Laenen and M. Spira, \np{511}{523}{98}.

\item \label{KS}
N. Kidonakis and G. Sterman, \pl{387}{867}{96}, \np{505}{321}{97}.

\item \label{KOS}
N. Kidonakis, G. Oderda and G. Sterman, \np{531}{365}{98}.

\item \label{LM}
E.\ Laenen and S.\ Moch, \pr{59}{034027}{99}.

\item \label{LOS}
E.\ Laenen, G. Oderda and G. Sterman, \pl{438}{173}{98}.

\item \label{CMNg}
S.\ Catani, M.L.\ Mangano and P.\ Nason, \jhep{07}{024}{98}.

\item \label{CMNOV}
S.\ Catani, M.L.\ Mangano, P.\ Nason, C.\ Oleari and W. Vogelsang,
\jhep{03}{025}{99}.

\item \label{kow}
N.~Kidonakis and J.F.~Owens, preprint FSU-HEP-991216 (hep-ph/9912388).

\item \label{cacciari}
M. Cacciari, preprint CERN-TH/99-312 (hep-ph/9910412).

\item \label{kdd}
N.~Kidonakis and V.~Del Duca, preprint FSU-HEP-991123 (hep-ph/9911460).

\item \label{tqll1}
E.~Laenen, J. Smith and W.L. van Neerven, \np{369}{543}{92}, \pl{321}{254}{94}. 

\item \label{tqll2}
E. Berger and H. Contopanagos, \pr{54}{3085}{96},  \pr{57}{253}{98}.

\item \label{tqll3}
S.\ Catani, M.L.\ Mangano, P.\ Nason and L.\ Trentadue, \pl{378}{329}{96}.

\item \label{tqll4}
N.\ Kidonakis, preprint EDINBURGH-99-4 (hep-ph/9904507).

\end{enumerate}

\end{document}